\documentclass{IEEEtaes}
\usepackage[stable]{footmisc}   
\usepackage{color,array,amsthm, soul}
\usepackage{graphicx}
\usepackage{float}
\usepackage{comment}
\usepackage{subfiles}

\usepackage{xcolor}
\usepackage[most]{tcolorbox}

\usepackage{colortbl}
\usepackage{xcolor}
\definecolor{headergray}{gray}{0.88}
\definecolor{lightgray}{gray}{0.94}
\definecolor{ourblue}{RGB}{214,228,245}  

\jvol{XX}
\jnum{XX}
\jmonth{XXXXX}
\paper{1234567}
\pubyear{2025}
\doiinfo{TAES.2025.Doi Number}

\usepackage{caption}
\usepackage{subcaption}

\usepackage{hyperref, cite} 
\usepackage{amssymb,amsfonts}
\usepackage{bm}
\usepackage{algorithmic}
\usepackage{graphicx}
\usepackage{textcomp}
\usepackage{color}  
\usepackage{booktabs}
\usepackage{caption}

\usepackage{footnote} 
\makesavenoteenv{table*} 

\usepackage[dvipsnames]{xcolor}

\newcommand{\helen}[1]{\textcolor{Orchid}{[{\em Helen}: #1]}}

\usepackage{amsmath}

\newcommand{\cp}[1]{\ifmmode {\mathcal{#1}}\else ${\mathcal{#1}}$\fi}
\usepackage{array}
\newcolumntype{P}[1]{>{\centering\arraybackslash}p{#1}}

\newcommand{\bA}{\mathbf{A}} 
\newcommand{\bB}{\mathbf{B}} 
\newcommand{\bC}{\mathbf{C}} 
\newcommand{\bD}{\mathbf{D}} 
 
\newcommand{\bE}{\mathbf{E}} 
\newcommand{\bF}{\mathbf{F}} 
\newcommand{\bG}{\mathbf{G}} 
\newcommand{\bH}{\mathbf{H}} 
\newcommand{\bI}{\mathbf{I}} 
\newcommand{\bJ}{\mathbf{J}}

\newcommand{\bM}{\mathbf{M}} 
\newcommand{\bN}{\mathbf{N}} 
\newcommand{\bP}{\mathbf{P}} 
 
\newcommand{\bR}{\mathbf{R}}

\newcommand{\bU}{\mathbf{U}} 
 
\newcommand{\bW}{\mathbf{W}} 
\newcommand{\bX}{\mathbf{X}} 
\newcommand{\bY}{\mathbf{Y}} 
\newcommand{\bZ}{\mathbf{Z}} 
\newcommand{\ba}{\mathbf{a}} 
\newcommand{\bb}{\mathbf{b}}

\newcommand{\bp}{\mathbf{p}} 
\newcommand{\be}{\mathbf{e}}

\newcommand{\by}{\boldsymbol{y}} 
 
\newcommand{\bt}{\mathbf{t}} 
\newcommand{\bu}{\mathbf{u}} 
\newcommand{\bv}{\mathbf{v}} 
\newcommand{\bx}{\mathbf{x}}

\newcommand{\bz}{\mathbf{z}}

\newcommand{\bvarepsilon}{\boldsymbol{\varepsilon}}

\newcommand{\bPhi}{\boldsymbol{\Phi}}
\newcommand{\brho}{\boldsymbol{\rho}}
\newcommand{\bomega}{\boldsymbol{\omega}}

\newcommand{\bGamma}{\boldsymbol{\Gamma}}

\newcommand{\btheta}{\boldsymbol{\theta}}

\newcommand{\bOmega}{\boldsymbol{\Omega}}

\newcommand{\bSigma}{\boldsymbol{\Sigma}}

\usepackage{nccmath}
\def\cred{\textcolor{red}}

\definecolor{darkgreen}{rgb}{0., 0.4, 0.}
\definecolor{amber}{rgb}{1.0, 0.49, 0.0}
\definecolor{orange}{rgb}{1.0, 0.4, 0.0}

\usepackage{xcolor}
\usepackage{amsmath,amsfonts}

\usepackage{algorithm2e}
\usepackage{mathtools}
\let\labelindent\relax
\usepackage{enumitem}
\usepackage{cleveref}
\usepackage{multirow}
\usepackage{array,cuted} 

\usepackage[toc,acronyms,nopostdot,nogroupskip, sort=use, nonumberlist]{glossaries-extra} 
\MFUhyphentrue 
\glssetcategoryattribute{general}{glossdesc}{firstuc}
\glssetcategoryattribute{abbreviation}{glossdesc}{title}

\makeatletter
\newglossarystyle{long-initcapsdesc}{%
  \setglossarystyle{long}%
    \renewcommand{\glsnamefont}[1]{\MakeUppercase{##1}}%
    \renewcommand{\glossentry}[2]{%
    \glsentryitem{##1}\glstarget{##1}{\glossentryname{##1}} &
    \protected@edef\thisdesc{\glsentrydesc{##1}}%
    \xcapitalisewords{\thisdesc}\glspostdescription\space ##2\tabularnewline
  }%
}
\makeatother

\makeglossaries

\usepackage{ragged2e}

\usepackage{amsthm}
\newtheoremstyle{boldsmallcaps}
  {\topsep}                       
  {\topsep}                       
  {\itshape}                      
  {}                              
  {\bfseries\scshape}             
  {.}                             
  { }                             
  {}                              

\theoremstyle{boldsmallcaps} 
\newtheorem{remark}{Remark}
\newtheorem{definition}{Definition}

\begin{document}
\newcommand*{\transpose}{\mathsf{T}}
\newcommand{\blockj}[1]{\text{block}_{#1}} 
\title{Cooperative Differential GNSS Positioning: Estimators and Bounds}
%
%
%
%
%
\author{Helena Calatrava}
\member{Student Member, IEEE}
\affil{Northeastern University, Boston, MA 02115, USA} 
\author{Daniel Medina}
\member{\textcolor{Black}{Senior Member, IEEE}}
\affil{German Aerospace Center (DLR), Neustrelitz, Germany} 
%
%
\author{Pau Closas}
\member{Senior Member, IEEE}
\affil{Northeastern University, Boston, MA 02115, USA} 
%
%
\receiveddate{Manuscript received XXXXX 00, 0000; revised XXXXX 00, 0000; accepted XXXXX 00, 0000. \\
This work was partially supported by the National Science Foundation under Awards 1845833, 2326559, and 2530870.}

\corresp{{\itshape (Corresponding author: Helena Calatrava)}.}

\authoraddress{Helena Calatrava and Pau Closas are with Dept.~of Electrical and Computer Eng., Northeastern University, Boston, MA (USA). E-mail: {\tt\small\{calatrava.h, closas\}@northeastern.edu}. Daniel Medina is with the Institute of Communications and Navigation, German Aerospace Center (DLR), Neustrelitz, Germany. E-mail: {\tt\small daniel.ariasmedina@dlr.de}.}

%
%
%
\markboth{CALATRAVA ET AL.}{ESTIMATORS AND BOUNDS FOR COOPERATIVE DIFFERENTIAL GNSS POSITIONING}
\maketitle
%
%
\begin{abstract}
In Differential GNSS (DGNSS) positioning, differencing measurements between a user and a reference station suppresses common-mode errors but also introduces reference-station noise, which fundamentally limits accuracy. This limitation is minor for high-grade stations but becomes significant when using reference infrastructure of mixed quality. This paper investigates how large-scale user cooperation can mitigate the impact of reference-station noise in conventional (non-cooperative) DGNSS systems. We develop a unified estimation framework for cooperative DGNSS (C-DGNSS) and cooperative real-time kinematic (C-RTK) positioning, and derive parameterized expressions for their Fisher information matrices as functions of network size, satellite geometry, and reference-station noise. This formulation enables theoretical analysis of estimation performance, identifying regimes where cooperation asymptotically restores the accuracy of DGNSS with an ideal (noise-free) reference. Simulations validate these theoretical findings.
\end{abstract}
%
%
\newabbreviation{CPC}{CPC}{central processing center}
\newabbreviation[shortplural={GDOPs}]{GDOP}{GDOP}{geometric dilution of precision}
\newabbreviation{FIM}{FIM}{Fisher information matrix}
\newabbreviation{GB}{GB}{global bound}
\newabbreviation{JPA}{JPA}{joint position and attitude}
\newabbreviation{O-C}{O-C}{observed-minus-computed}
\newabbreviation{SPP}{SPP}{standard point positioning}
\newabbreviation{PVT}{PVT}{position, velocity, and time}
\newabbreviation{PPP}{PPP}{precise point positioning}
\newabbreviation{DGNSS}{DGNSS}{differential GNSS}
\newabbreviation{RTK}{RTK}{real-time kinematic}
\newabbreviation{WADGNSS}{WADGNSS}{wide-area differential GNSS}
\newabbreviation{NLOS}{NLOS}{non-line-of-sight}
\newabbreviation{CP}{CP}{cooperative positioning}
\newabbreviation{MDS}{MDS}{multidimensional scaling}
\newabbreviation{CDOP}{CDOP}{cooperative dilution of precision}
\newabbreviation{V2X}{V2X}{vehicle-to-everything}
\newabbreviation{C-DGNSS}{C-DGNSS}{cooperative DGNSS}
\newabbreviation{C-RTK}{C-RTK}{cooperative RTK}
\newabbreviation{MLE}{MLE}{maximum likelihood estimator}
\newabbreviation{MAP}{MAP}{maximum a posteriori}
\newabbreviation{CRB}{CRB}{Cramér-Rao bound}
\newabbreviation{ecdf}{ECDF}{empirical cumulative distribution function}
\newabbreviation{RMSE}{RMSE}{root mean square error}
\newabbreviation{RL}{RL}{reinforcement learning}
\newabbreviation{RJDM}{RJDM}{radar jamming decision-making}
\newabbreviation{LORO}{LORO}{lobe-on-receive-only}
\newabbreviation{RPIP}{RPIP}{random pulse initial phases}
%
%
\newabbreviation{CPI}{CPI}{coherent processing interval}
\newabbreviation{IC}{IC}{interference cancellation}
\newabbreviation{MUSIC}{MUSIC}{multiple signal classification}
\newabbreviation{SAM}{SAM}{swept amplitude-modulation}
\newabbreviation{AM}{AM}{swept amplitude-modulation}
\newabbreviation{NP}{NP}{Neyman-Pearson}
\newabbreviation{DBS}{DBS}{Doppler beam sharpening}
\newabbreviation{MTT}{MTT}{multiple target tracking}
\newabbreviation{LFM}{LFM}{linear frequency-modulated}
\newabbreviation{LOS}{LOS}{line-of-sight}
\newabbreviation{SPJ}{SPJ}{self-protection jamming}
\newabbreviation{SOJ}{SOJ}{stand-off jamming}
\newabbreviation{EJ}{EJ}{escort jamming}
\newabbreviation{JRC}{JRC}{joint radar-communication}
\newabbreviation{WLS}{WLS}{weighted least squares}
\newabbreviation{RSS}{RSS}{received signal strength}
\newabbreviation{TOA}{TOA}{time of arrival}
\newabbreviation{ERP}{ERP}{effective radiated power}
\newabbreviation{LTI}{LTI}{linear time-invariant}
\newabbreviation[shortplural={KFs}, plural={Kalman filters}]{KF}{KF}{Kalman filter}
\newabbreviation[shortplural={DePINs}, plural={decentralized physical infrastructure networks}]{DePIN}{DePIN}{decentralized physical infrastructure network}

\newabbreviation{MSB}{MSB}{McAulay-Seidman}

\newabbreviation{UWB}{UWB}{ultra-wideband}

\newabbreviation[shortplural={CORSs}, plural={continuously operating reference stations}]{CORS}{CORS}{continuously operating reference station}

\newabbreviation[shortplural={EKFs}, plural={extended Kalman filters}]{EKF}{EKF}{extended Kalman filter}
\newabbreviation[shortplural={PFs}, plural={particle filters}]{PF}{PF}{particle filter}
\newabbreviation[shortplural={UKFs}, plural={unscented Kalman filters}]{UKF}{UKF}{unscented Kalman filter}
\newabbreviation[shortplural={GSFs}, plural={Gaussian sum filters}]{GSF}{GSF}{Gaussian sum filter}
\newabbreviation[shortplural={DOFs}, plural={degrees of freedom}]{DOF}{DOF}{degree of freedom}
\newabbreviation[shortplural={MMFTs}, plural={micro-motion false targets}]{MMFT}{MMFT}{micro-motion false target}
\newabbreviation[shortplural={TFTs}, plural={translational false targets}]{TFT}{TFT}{translational false target}
\newabbreviation[shortplural={PRIs}, plural={pulse repetition interval}]{PRI}{PRI}{pulse repetition interval}
\newabbreviation[shortplural={PTs}, plural={physical targets}]{PT}{PT}{physical target}
\newabbreviation{FFT}{FFT}{Fast Fourier Transform}
\newabbreviation{RCS}{RCS}{radar cross-section}
\newabbreviation{CS}{CS}{compressed sensing}
\newabbreviation{AI}{AI}{artificial intelligence}
\newabbreviation{EW}{EW}{electronic warfare}
\newabbreviation{CW}{CW}{continuous-wave}
\newabbreviation{BSS}{BSS}{blind signal synthesis}
\newabbreviation{IF}{IF}{intermediate frequency}
\newabbreviation{LiDAR}{LiDAR}{light detection and ranging}
\newabbreviation{INS}{INS}{inertial navigation system}

\newabbreviation[shortplural={TOIs}, plural={targets of interest}]{TOI}{TOI}{target \MakeLowercase{o}f interest}
\newabbreviation[shortplural={ECMs}, plural={electronic countermeasures}]{ECM}{ECM}{electronic countermeasure}
\newabbreviation[shortplural={ECCMs}, plural={electronic counter-countermeasures}]{ECCM}{ECCM}{electronic counter-countermeasure}
\newabbreviation[shortplural={FTs}, plural={false targets}]{FT}{FT}{false target}
\newabbreviation[shortplural={PDs}, plural={pulse dopplers}]{PD}{PD}{pulse doppler}
\newabbreviation[shortplural={FTGs}, plural={false target generator}]{FTG}{FTG}{false target generator}
\newabbreviation[shortplural={FDAs}, plural={frequency diverse arrays}]{FDA}{FDA}{frequency diverse array}
\newabbreviation[shortplural={SARs}, plural={synthetic aperture radars}]{SAR}{SAR}{synthetic aperture radar}
\newabbreviation[shortplural={DRFMs}, plural={digital radio-frequency memories}]{DRFM}{DRFM}{digital radio frequency memory}
\newabbreviation[shortplural={RGPOs}, plural={range gate pull-offs}]{RGPO}{RGPO}{range gate pull-off}
\newabbreviation[shortplural={DDs}, plural={double differences}]{DD}{DD}{double difference}
\newabbreviation[shortplural={SDs}, plural={single differences}]{SD}{SD}{single difference}
\newabbreviation[shortplural={RGPIs}, plural={range gate pull-ins}]{RGPI}{RGPI}{range gate pull-in}
\newabbreviation[shortplural={VGPOs}, plural={velocity gate pull-offs}]{VGPO}{VGPO}{velocity gate pull-off}
\newabbreviation[shortplural={VGPI}s, plural={velocity gate pull-ins}]{VGPI}{VGPI}{velocity gate pull-in}
\newabbreviation[shortplural={RVGPOs}, plural={range-velocity gate pull-offs}]{RVGPO}{RVGPO}{range-velocity gate pull-off}
\newabbreviation[shortplural={RVGPIs}, plural={range-velocity gate pull-ins}]{RVGPI}{RVGPI}{range-velocity gate pull-in}
\newabbreviation[shortplural={ISRJs}, plural={interrupted-sampling repeater jammings}]{ISRJ}{ISRJ}{interrupted-sampling repeater jamming}
\newabbreviation[shortplural={MHTs}, plural={multiple hypothesis trackings}]{MHT}{MHT}{multiple hypothesis tracking}
\newabbreviation[shortplural={RNNs}, plural={recurrent neural networks}]{RNN}{RNN}{recurrent neural network}
\newabbreviation[shortplural={SJNRs}, plural={Signal-to-Jammer Noise Ratios}]{SJNR}{SJNR}{Signal-to-Jammer Noise Ratio}
\newabbreviation[shortplural={CNNs}, plural={convolutional neural networks}]{CNN}{CNN}{convolutional neural network}
\newabbreviation[shortplural={LSTMs}, plural={long short-term memories}]{LSTM}{LSTM}{long short-term memory}
\newabbreviation[shortplural={TDOAs}, plural={time differences of arrival}]{TDOA}{TDOA}{time difference of arrival}
\newabbreviation[shortplural={RFSs}, plural={random finite sets}]{RFS}{RFS}{random finite set}
\newabbreviation[shortplural={SNRs}, plural={signal-to-noise ratios}]{SNR}{SNR}{signal-to-noise ratio}
\newabbreviation[shortplural={SJRs}, plural={signal-to-jammer ratios}]{SJR}{SJR}{signal-to-jammer ratio}
\newabbreviation[shortplural={OFDMs}, plural={orthogonal frequency-division multiplexings}]{OFDM}{OFDM}{orthogonal frequency-division multiplexing}
\newabbreviation[shortplural={PRFs}, plural={pulse repetition frequencies}]{PRF}{PRF}{pulse repetition frequency}
\newabbreviation[shortplural={SIMOs}, plural={single-input multiple-outputs}]{SIMO}{SIMO}{single-input multiple-output}
\newabbreviation[shortplural={MIMOs}, plural={multiple-input multiple-outputs}]{MIMO}{MIMO}{multiple-input multiple-output}
\newabbreviation[shortplural={GLRTs}, plural={generalized likelihood ratio tests}]{GLRT}{GLRT}{generalized likelihood ratio test}
\newabbreviation[shortplural={AGCs}, plural={automatic gain controls}]{AGC}{AGC}{automatic gain control}
\newabbreviation[shortplural={MLs}, plural={machine learnings}]{ML}{ML}{machine learning}
\newabbreviation[shortplural={UAVs}, plural={unmanned aerial vehicles}]{UAV}{UAV}{unmanned aerial vehicle}
\newabbreviation[shortplural={GNSSs}, plural={global navigation satellite systems}]{GNSS}{GNSS}{global navigation satellite systems}

%

\renewcommand{\glsnamefont}[1]{\makefirstuc{#1}}
%

\vspace{-0.2cm}

\section{Introduction}


%

\Gls{GNSS} remains the cornerstone of outdoor navigation, providing globally accessible, reliable, and drift-free positioning capabilities~\cite{Groves_principles,morton2021position, egea2022gnss}.
However, standalone \gls{GNSS} positioning offers only meter-level accuracy due to the limited precision of code measurements and residual biases arising from imperfect orbit and atmospheric modeling~\cite[Ch.~21]{teunissen_book}.

While meter-level accuracy is sufficient for most consumer applications, emerging domains such as intelligent transportation systems and \gls{V2X} communications~\cite{Williams_SAE22,MENDES2025100878} demand \textit{centimeter-level precision}~\cite{hassan2021review,an2023array}.
\Gls{PPP}~\cite[Ch. 25]{teunissen_book} offers a means to achieve such accuracy by incorporating precise orbit and clock corrections; however, its long convergence time limits its suitability for real-time applications~\cite{wabbena2005ppp,hou2023recent}.

\begin{table*}[t]
\centering
\caption{Overview of representative GNSS cooperative positioning approaches by measurement source; see~\cite{jin2024survey} for a broad survey. We focus on the C-DGNSS and C-RTK methodologies (blue row), which rely on GNSS measurements differenced with those of a surveyed base station, without using auxiliary sensing.
}
\label{tab:cp_taxonomy}
\begin{tabular}{@{\extracolsep{\fill}}p{5.5cm} p{6.3cm} p{4.3cm}}
\toprule
\textbf{Measurements} & \textbf{Cooperation mechanism} & \textbf{Georeferenced?} \\
\midrule
GNSS-only~\cite{6691955,minetto2019information,sfsf, minetto2022dgnss,coop_gnss_competencia,rs16152755,minetto2022dgnss}
& Inter-user constraint formation via GNSS differencing (DGNSS-CP) to cancel common errors
& No, unless GNSS measurements are reused \\
\midrule
GNSS + inter-user sensing (e.g., ranges, coordinate differences)~\cite{7112186,jordi_multi,schaper2024multi,penna2010cramer}
& Fusion of absolute GNSS information and inter-user constraints from auxiliary sensing
& Yes, through absolute (undifferenced) GNSS observations \\
\midrule
\rowcolor{ourblue}
GNSS + reference information (e.g., probabilistic prior, surveyed station)~\cite{lee2012rfid,du2008next,rezaei2007kalman,zhang2019rectification,zhang20203d,calatrava_massive_2023,calatrava2024collaborative,medina2023collaborative,calatrava_cdgnss_2025}
& User-anchor constraint formation to cancel common errors
& Yes, typically through the reference entity \\
\bottomrule
\end{tabular}
\end{table*}


Real-time precise positioning is enabled by \gls{DGNSS}~\cite[Ch.~26]{teunissen_book}. In this architecture, professionally maintained \glspl{CORS} at surveyed locations~\cite[Ch.~6]{ogaja2022gnss} broadcast corrections that mitigate errors common to both the reference station and nearby user receivers~\cite{kim2024expanding}. Code-based DGNSS implementations rely exclusively on pseudorange measurements, typically achieving decimeter-level accuracy. Further incorporating carrier-phase measurements, \gls{RTK} resolves integer ambiguities and enables centimeter-level precision~\cite[Ch.~2]{medina2022robust}.


\begin{figure}[t]
\centering
\centerline{\includegraphics[width=.95\columnwidth]{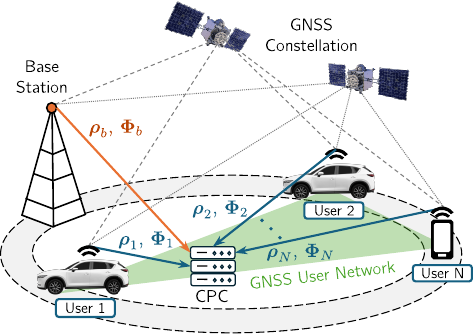}}
\caption{Overview of the cooperative DGNSS architecture considered in this work. Nearby users $r\in{1,\ldots,N}$ transmit code ($\brho_r$) and carrier-phase ($\bPhi_r$) measurements to a CPC, which jointly processes them with base-station DGNSS corrections. Cooperation is centralized, base-station–anchored, and does not involve inter-user ranging.}
\vspace{-.8cm}
\label{fig:intro_cdgnss}
\end{figure}

Remarkably, most reference networks are proprietary and managed by national or private operators, prompting growing efforts to make precise positioning more widely accessible~\cite{elneser2023communitycors}.
This includes the emergence of low-cost GNSS receivers supporting RTK- and PPP-level capabilities~\cite{stempfhuber2012precise,robustelli2023low,li2023performance}, efforts by commercial providers to broaden access to precise correction services~\cite{horton2023geodnet}, and public services offering openly accessible real-time PPP corrections. The latter includes CNES products~\cite{zhao2020investigation} and the Galileo High Accuracy Service (HAS)~\cite{fernandez2022galileo}.

Within this ongoing \textit{democratization of precise positioning}, we identify achieving CORS-grade performance using low-cost reference infrastructure as a key research objective. While low-cost receivers reduce deployment costs, their limited hardware quality degrades the accuracy of \gls{DGNSS} and \gls{RTK} corrections.
One approach to address this challenge is to leverage the growing availability of data-sharing infrastructures and employ \textit{\gls{CP}} techniques~\cite{jin2024survey} (see Sec.~\ref{sec:related_work} for a brief review).
When multiple receivers rely on a common reference station for \gls{DGNSS} corrections, their differential observations become statistically correlated through the shared reference noise. Accounting for these correlations, which are particularly pronounced when low-cost reference stations are used, enhances positioning accuracy.
%
%
%

In this work, we adopt an estimation-theoretic perspective to address the following questions:
\textit{Can large-scale user cooperation compensate for the reference-station noise propagated through the differencing operation in standard (non-cooperative) DGNSS? If so, under which conditions, and to what extent?}
The main contributions of this paper, relative to our prior conference works in which the potential of this cooperative positioning approach was preliminarily explored~\cite{medina2023collaborative,calatrava_cdgnss_2025}, are:
\begin{itemize}
    \item We present a unified estimation framework for \textit{cooperative differential GNSS} that integrates both code- and carrier-phase-based techniques and accounts for a reference station with arbitrary noise variance.
    \item  We provide a comprehensive theoretical analysis of the asymptotic performance of \gls{C-DGNSS}~\cite{calatrava_cdgnss_2025} and \gls{C-RTK}~\cite{medina2023collaborative} based on the \gls{FIM}, and as a function of receiver count, satellite visibility, and measurement variance.

    \item We validate the proposed framework through simulations representative of multi-user network scenarios, such as the one illustrated in Fig.~\ref{fig:intro_cdgnss}.
\end{itemize}

The remainder of this paper is organized as follows:
Sec.~\ref{sec:related_work} reviews related work;
Sec.~\ref{sec:method} formalizes the \gls{C-DGNSS} and \gls{C-RTK} methodologies;
Sec.~\ref{sec:bounds} analyzes their theoretical performance limits;
Sec.~\ref{sec:simulations} presents the simulation results;
and Sec.~\ref{sec:conclusion} concludes the paper.

\section{Background on GNSS Cooperative Positioning}\label{sec:related_work}
This section outlines representative approaches to \gls{GNSS} \gls{CP} as summarized in Table~\ref{tab:cp_taxonomy}, and introduces the cooperative architecture considered in this work, as illustrated in Fig.~\ref{fig:intro_cdgnss}.

Interest in \gls{GNSS} \gls{CP} emerged in the early 2000s~\cite{heinrichs2004hybrid}, driven by advances in wireless sensor networks and short-range communication technologies that enabled information exchange among receivers~\cite{buehrer2018collaborative}. Notably, CP was originally developed for mass-market and indoor environments, where external positioning infrastructure was scarce and devices had to rely on inter-receiver cooperation to overcome low SNR and limited transmitter availability~\cite{wymeersch2009cooperative,van2015least}.
Since the late 2010s, renewed attention has been fueled by the availability of raw GNSS measurements on mass-market devices~\cite{gogoi2019cooperative,icking2022optimal,weng2020new,minetto2022dgnss} and the rise of \gls{V2X} communications~\cite{minetto_gnssOnly,zhuang2021cooperative}.

%
A well-established approach is \gls{DGNSS}-\gls{CP}~\cite{6691955,minetto2019information,sfsf}, which operates without any reference by differencing the measurements of multiple receivers to cancel common-mode errors and infer inter-agent ranges~\cite[\S IV.C]{jin2024survey}. These cooperative ranges can then be incorporated as additional constraints in the PVT solver~\cite{minetto2022dgnss,coop_gnss_competencia}, thereby improving absolute positioning for users with limited satellite visibility~\cite{rs16152755,minetto2022dgnss}.


On the other hand, hybrid \gls{GNSS} \gls{CP} directly processes raw GNSS measurements while integrating relative information from independent ranging technologies~\cite{7112186,jordi_multi}. This includes inter-receiver ranges obtained from \gls{RSS} or \gls{TOA} measurements, with the former being highly sensitive to the environment. When implemented with \gls{UWB} technology, \gls{TOA} ranging can achieve centimeter-level accuracy \cite{dardari2015indoor}. Remarkably, some works use vision-based or other relative-pose sensing modalities that provide vector-valued constraints, substantially strengthening network observability~\cite{schaper2024multi}.
Theoretical analysis of hybrid architectures was first formalized in~\cite{penna2010cramer}, and~\cite{7112186} introduced the CDOP metric to demonstrate the benefits of increased network connectivity and geometric diversity.

Some methods anchor network positioning to a global frame through reference points (also referred to as network anchors) or position initialization for selected users~\cite{lee2012rfid,du2008next,rezaei2007kalman}.
%
%
Our previous work introduced a massive differentiation method~\cite{calatrava_massive_2023}, in which a target user forms differential measurements with neighboring receivers whose positions are modeled via probabilistic priors, allowing these neighbors to act as virtual references. Under the assumption that such priors are zero-mean and hence unbiased, the method achieves asymptotic \gls{DGNSS}/\gls{RTK}-level performance without requiring a physical base station~\cite{calatrava2024collaborative}.
%


\begin{figure*}[t]
\centering
\centerline{\includegraphics[width=.95\textwidth]{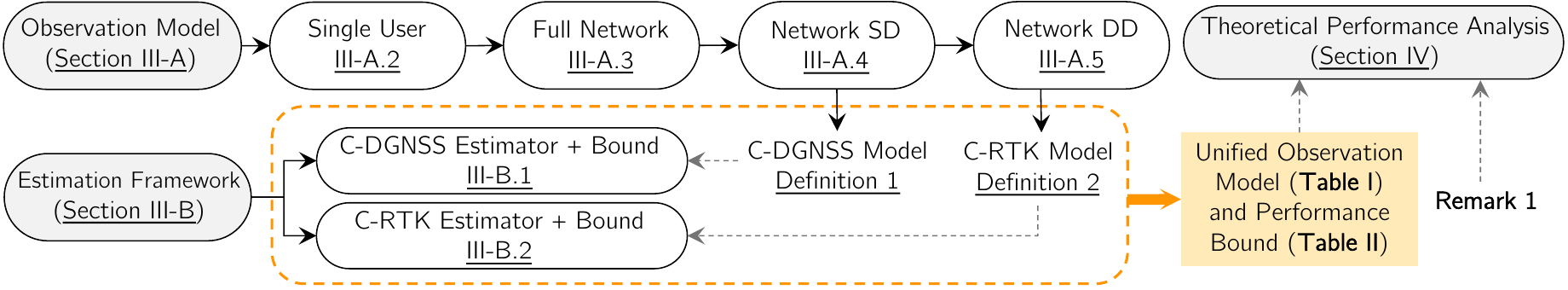}}
\caption{Overview of the derivation flow in Sec.~\ref{sec:method} leading to a unified observation model and performance bound for C-DGNSS and C-RTK, summarized in Tables~\ref{tab:crb_models} and~\ref{tab:crb_model_summary}. The resulting unified \gls{FIM} formulation enables the theoretical analysis in Sec.~\ref{sec:bounds}, which constitutes the main contribution of this work. Solid arrows indicate section flow, while dashed arrows denote result reuse across derivations.}
\vspace{-.3cm}
\label{fig:organization}
\end{figure*}

\textbf{Cooperative Differential GNSS Architecture:} In this work, we consider a \gls{C-DGNSS}~\cite{calatrava_cdgnss_2025} and \gls{C-RTK}~\cite{medina2023collaborative} architecture, where the network comprises one surveyed reference station (indexed by $b$) and $N$ user receivers (indexed by $r \in \{1,\ldots,N\}$). The reference station observes a set of $K$ satellites (indexed by $s \in \{p,1,\ldots,K{-}1\}$), where $p$ denotes the pivot satellite used for differencing. Each user receiver observes all or a subset of these satellites depending on visibility. Cooperation is realized through \emph{centralized processing}: all receivers (and the reference station) share their raw code and carrier-phase measurements with a common central processing center (CPC), which performs joint state estimation. Importantly, there is \emph{no inter-user ranging}; all differential measurements are formed between each user and the reference station, providing a shared differencing anchor and enabling explicit modeling of inter-measurement correlations across the network.


Addressing multipath and \gls{NLOS} effects in \gls{CP} remains challenging, as these depend strongly on each user’s local geometry and environment, thus leading to largely uncorrelated measurements~\cite[\S IV.B]{jin2024survey}. Map-aided~\cite{zhang20203d} and machine-learning-based~\cite{li2023machine} methods show promise, with some exploiting spatial correlations among users in similar geometries~\cite{icking2022optimal} or benefiting from sensor integration~\cite{tanwar2018decentralized}.
This work instead targets the mitigation of atmospheric effects that are common to all receivers and to the low-cost reference station operating under short-baseline conditions (typically $<$10 km). Multipath, \gls{NLOS}, privacy preservation~\cite{hernandez2020privacy}, and time-synchronization issues~\cite{minetto2022dgnss} fall outside the scope of this paper.
Robustness to outliers in the \gls{C-DGNSS} framework is investigated in~\cite{calatrava2025towards}.

\section{Cooperative DGNSS and RTK Methodologies}\label{sec:method}
%
%

This section presents a unified estimation framework for the \gls{C-DGNSS} and \gls{C-RTK} methodologies under a reference station with arbitrary noise variance, covering their linearized observation models (Sec.~\ref{sec:method}-\ref{sec:observation_model}) as well as the associated estimation problem and performance bounds (Sec.~\ref{sec:method}-\ref{sec:estimation_problem_bounds}). The reader should refer to Fig.~\ref{fig:organization} for an overview of this section.
%



\textbf{Notation:} Bold lowercase and uppercase letters denote vector- and matrix-valued quantities, respectively, while regular lowercase letters denote scalars.
The all-ones vector is $\underline{1}_{p} \in \mathbb{R}^{p}$; $\bI_p$ is the $p \times p$ identity matrix; the matrices $\mathbf{1}_{p,q}, \mathbf{0}_{p,q} \in \mathbb{R}^{p \times q}$ contain ones and zeros, respectively, with the shorthand $\mathbf{1}_{p} \triangleq \mathbf{1}_{p,p}$ and $\mathbf{0}_{p} \triangleq \mathbf{0}_{p,p}$; $\otimes$ is the Kronecker product; and $\nabla_{\mathbf{x}}(\cdot)$ denotes the gradient w.r.t. $\mathbf{x}$.
%
%

\vspace{-.35cm}

\subsection{Observation Models}\label{sec:observation_model}
In this section, we formulate the \gls{C-DGNSS} and \gls{C-RTK} methodologies within a unified Gaussian model:
\begin{equation}\label{eq:unified_model_1}
    \by|
    \btheta \sim \mathcal{N}(\mathbf{m}(\btheta),\,\bSigma),\ \btheta^\transpose = [\bomega^\transpose,\ \bz^\transpose],
\end{equation}
where \(\by \in \mathbb{R}^{M_y}\) represents the measurement vector, and \(\btheta \in \mathbb{R}^{M_\theta}\) is the unknown (deterministic) parameter vector, composed of continuous parameters \(\bomega \in \mathbb{R}^{M_\omega}\) and integer parameters \(\bz \in \mathbb{Z}^{M_z}\), such that \(M_\theta = M_\omega + M_z\). 
The function \(\mathbf{m}(\btheta) \in \mathbb{R}^{M_y}\) and \(\bSigma \in \mathbb{R}^{M_y\times M_y}\) are the measurement model mean and covariance.

%
%
We adopt the following simplifying assumptions, until otherwise stated in Sec.~\ref{sec:bounds}-\ref{sec:network_clustering}:
\begin{enumerate}
    \item[A1)] All users \(r \in \{1,\ldots,N\}\) and the base station track the same set of \(K\) satellites \(s \in \{p,1,\ldots,K{-}1\}\).
    \item[A2)] The stochastic measurement models share a common structure across users.
    \item[A3)] Users are in close proximity, implying identical satellite–receiver \gls{LOS} vectors across the network.
\end{enumerate}

We next derive the explicit forms of $\btheta$, $\mathbf{m}(\btheta)$, and $\bSigma$ for the \gls{C-DGNSS} and \gls{C-RTK} models, as summarized in Table~\ref{tab:crb_models}. 
The derivation follows four steps, each detailed in Secs.~\ref{sec:method}-\ref{sec:observation_model}.\ref{sec:step1} through \ref{sec:method}-\ref{sec:observation_model}.\ref{sec:step4}: 
\textit{(1)} definition and linearization of the single-user observation model; \textit{(2)} introducing a compact notation to stack all network observations; 
\textit{(3)} applying the \gls{SD} operator with respect to the reference station to obtain the \gls{C-DGNSS} model (Definition~\ref{def:definition_1}); and 
\textit{(4)} extending to the \gls{DD} case by selecting a reference satellite, yielding the \gls{C-RTK} model (Definition~\ref{def:definition_2}).
Note that expanded formulations are provided in Sec.~S1 of the Supplemental Material.

\subsubsection{Definition of Cooperative Operators}\label{sec:operators}

We first introduce three operators that compactly represent the stacking and differencing operations used in the subsequent derivations.
Throughout this section, let $m$ denote the dimension of the vector to which stacking or differencing has been applied. In practice, $m$ is either the number of satellites (for measurement differencing) or the state-vector dimension.

\textit{a) Cooperative Stacking Operator:} To stack the observations from the full network, we define
\begin{equation}\label{eq:coop_op_vector}
\tilde{\bu} \triangleq [\, \bu_b^\top, \; \bu_1^\top, \; \dots, \; \bu_N^\top \,]
^\top \in \mathbb{R}^{m(N+1)},
\end{equation}
with $\bu_i \in \mathbb{R}^m$ denoting an arbitrary vector associated with either the reference station ($i = b$) or a user receiver ($i \in \{1, \ldots, N\}$).

\textit{b) Cooperative SD Operator:} With respect to the reference station~$b$, we define
\begin{equation}\label{eq:general_sd_vector}
\tilde{\bu}_b \triangleq \tilde{\bD}_b \tilde{\bu} = 
\begin{bmatrix}
\bu_{b1}^\top,\ \dots,\ \bu_{bN}^\top
\end{bmatrix}^\top \in \mathbb{R}^{mN},
\end{equation}
where \(\bu_{br} \triangleq \bu_r - \bu_b \in\mathbb{R}^m,\forall r \in \{1, \ldots, N\}\), denotes the per-user \gls{SD} quantity, and the \gls{SD} matrix has the form
\begin{equation}
\tilde{\bD}_b = \left[-\underline{1}_N \otimes \mathbf{I}_m,\ \mathbf{I}_N \otimes \mathbf{I}_m\right] \in\mathbb{R}^{mN\times m(N{+}1)}.
\end{equation}

\textit{c) Cooperative DD Operator:} With respect to a pivot satellite \(p\), we define
\begin{equation}\label{eq:operator_dd}
    \tilde{\bu}_b^p \triangleq \tilde{\bD}_p \tilde{\bu}_b = 
\begin{bmatrix}
(\bu_{b1}^p)^\top,\ \dots,\ (\bu_{bN}^p)^\top
\end{bmatrix}^\top \in \mathbb{R}^{(m-1)N},
\end{equation}
being $(\bu_{br}^p)^\transpose = (\bD_p\bu_{br})^\transpose \triangleq 
\begin{bmatrix}
u_{br}^{p1},\ \hdots,\ u_{br}^{p(m-1)}
\end{bmatrix}\in\mathbb{R}^{m{-}1}$ the per-user \gls{DD} vector, and $u_{br}^{ps} \triangleq u^s_{br} - u^p_{br} = u^s_{r} - u^p_{r} - (u^s_{b} - u^p_{b})\in\mathbb{R}$ the per-user per-satellite differentiation scalar.
The \gls{DD} matrices are defined as
\begin{equation}
\begin{split}
    \tilde{\bD}_p &= \mathbf{I}_N \otimes \bD_p\in\mathbb{R}^{(m{-}1)N\times mN},\\
    \bD_p &= [-\underline{1}_{m-1},\, \mathbf{I}_{m-1}]\in\mathbb{R}^{(m{-}1)\times m}.
\end{split}
\end{equation}
Note that the one-user DD matrix, $\bD_p$, has already been presented in the literature~\cite[Ch. 26]{teunissen_book}.

\begin{table*}[t]
\centering
\caption{Unified representation of parameters for each cooperative GNSS model, mapped to the Gaussian form in~\eqref{eq:unified_model_1}.}
\label{tab:crb_models}
\begin{tabular}{llllllllll}
\toprule
\textbf{Method} & \textbf{Model} & \(\by\) & \(M_y\) & \(\mathbf{m}(\btheta)\)& \(\bSigma\) & \(\bomega\) & \(M_\omega\) & \(\bz\) & \(M_z\) \\
\midrule
C-DGNSS          & \eqref{eq:cdgnss_model}        & \(\Delta\tilde{\brho}_b\) & \(KN\)        & \(\tilde{\bH}_N\btheta\)               & \(\tilde{\boldsymbol{\Sigma}}_{b,\rho}\)     & \(\delta\tilde{\bx}_b\) & \(4N\)       & \(\varnothing\) & \(0\) \\
C-RTK & \eqref{eq:crtk_general_model}   & \(\left[\Delta \tilde{\boldsymbol{\Phi}}_{b}^{{p\top}},\ \Delta \tilde{\boldsymbol{\rho}}_{b}^{{p\top}}\right]^\top\) & \(2(K{-}1)N\) & \([\tilde{\bB}\ \tilde{\bA}]\btheta\)  & \(\tilde{\boldsymbol{\Sigma}}_{b}^p\)   & \(\tilde{\bb}\)          & \(3N\)       & \(\tilde{\ba}\) & \(N(K{-}1)\) \\
\bottomrule
\vspace{-.75cm}
\end{tabular}
\end{table*}

\subsubsection{Single-User Model}\label{sec:step1}

At a given epoch, receiver~$r$ measures the pseudorange and carrier phase with respect to satellite~$s$ as
\begin{equation}\label{eq:pseudorange_meas}
\begin{split}
\rho_r^s &= \varrho_r^s + T_r^s + I_r^s + c(d t_r - d t^s) + \varepsilon_r^s \in \mathbb{R},\\
\Phi_r^s &= \varrho_r^s + T_r^s - I_r^s + c(d t_r - d t^s) + \lambda N_r^s + \epsilon_r^s \in \mathbb{R},
\end{split}
\end{equation}
where $\varrho_r^s \triangleq \|\bp^s - \bp_r\|$ denotes the true satellite–receiver range, with $\bp^s$ and $\bp_r$ being the satellite and receiver position vectors; $d t^s$ and $d t_r$ are the satellite and receiver clock offsets; $c$ is the speed of light; $\varepsilon_r^s$ and $\epsilon_r^s$ represent unmodeled errors (e.g., thermal noise, multipath, atmospheric residuals, ephemeris errors); $I_r^s$ and $T_r^s$ denote the ionospheric and tropospheric delays; $\lambda$ is the carrier wavelength; and $N_r^s$ is the integer ambiguity~\cite[Ch.~2]{medina2022robust}.
For simplicity, we omit frequency and time reference labels and neglect ephemeris errors.

For receiver $r$, the state vector collecting the unknown position and clock offset parameters is given by
\begin{equation}\label{eq:state_vector_linearization}
    \bx_r = [\bp_r^\top,\ c\,d t_r]^\top\in \mathbb{R}^4,\ \quad\delta\bx_r \triangleq \bx_r - \bx_{r,0},
\end{equation}
where $\mathbf{x}_{r,0}$ denotes the linearization (assumed reference) state used below. 
%


To obtain a locally linear and tractable model, the true range $\varrho_r^s$ is linearized with respect to $\bp_r$ by a first-order Taylor expansion around the initial estimate $\bp_{r,0}$:
\begin{equation}
\varrho_r^s \approx \|\bp^s - \bp_{r,0}\| - (\be_r^s)^\top \delta\bp_r,
\end{equation}
being $\delta\bp_r = \bp_r - \bp_{r,0}$ the position linearization increment, and $\be_r^s = -(\bp^s - \bp_{r,0}) / \|\bp^s - \bp_{r,0}\|$ the satellite–receiver \gls{LOS} unit vector. For brevity, the dependence of $\be_r^s$ on $\bp_{r,0}$ is omitted in the notation.


%
Substituting the linearized range into~\eqref{eq:pseudorange_meas} and expressing the result in \gls{O-C} form yield the following pseudorange and carrier-phase residuals for receiver~$r$ over the full set of $K$ visible satellites~\cite[Ch. 21]{teunissen_book}:
\begin{equation}\label{eq:model:approximation_single}
\begin{split}
\Delta\boldsymbol{\rho}_r &= \bH_r\,\delta\bx_r + c\,d\bt + \bm{T}_r + \bm{I}_r + \boldsymbol{\varepsilon}_r \in \mathbb{R}^K,\\
\Delta\boldsymbol{\Phi}_r &= \bH_r\,\delta\bx_r + c\,d\bt + \bm{T}_r - \bm{I}_r + \lambda\bN_r + \boldsymbol{\epsilon}_r \in \mathbb{R}^K,
\end{split}
\end{equation}
where $\bH_r = [\bE_r,\ \underline{1}_K] \in \mathbb{R}^{K\times4}$ is the observation model matrix, with $\bE_r = -\,[\be_r^1,\ldots,\be_r^K]^\top$ denoting the geometry matrix. 
Under Assumption~A3, \(\mathbf{E}_r = \mathbf{E}\) and \(\mathbf{H}_r = \mathbf{H}\) for all \(r\). The term \(c\,d\mathbf{t}\) includes only the satellite clock, as the receiver clock is included in the state.
%

\textcolor{black}{Note that the above single-frequency model can be extended to the multi-frequency case by stacking observations across frequencies and augmenting the state vector with the corresponding carrier-phase ambiguities.}

\subsubsection{Multi-User Model}\label{sec:step2}
Applying the stacking operator in~\eqref{eq:coop_op_vector} to~\eqref{eq:model:approximation_single} yields the full-network \gls{O-C} measurement equations:
\begin{equation}\label{eq:multi-user_obs}
\begin{split}
\Delta \tilde{\boldsymbol{\rho}} &=\tilde{\bH}_{N+1} \delta \tilde{\bx} +  c\,\tilde{d\bt} + \tilde{\bm{T}} + \tilde{\bm{I}} + \tilde{\boldsymbol{\varepsilon}} \in \mathbb{R}^{K(N+1)}, \\
\Delta \tilde{\boldsymbol{\Phi}} &= \tilde{\bH}_{N+1} \delta \tilde{\bx} + c\,\tilde{d\bt} + \tilde{\bm{T}} - \tilde{\bm{I}} + \lambda \tilde{\bN} + \tilde{\boldsymbol{\epsilon}} \in \mathbb{R}^{K(N+1)},
\end{split}
\end{equation}
with \(\tilde{\bH}_{N} = \bI_{N} \otimes \bH\).
For clarity, we explicitly write the following stacked quantities, based on~\eqref{eq:coop_op_vector}, as
\begin{align*}
\Delta\tilde{\boldsymbol{\rho}} &\triangleq
[\,\Delta\boldsymbol{\rho}_b^\top,\ \Delta\boldsymbol{\rho}_1^\top,\ \ldots,\ \Delta\boldsymbol{\rho}_N^\top\,]^\top\in\mathbb{R}^{K(N+1)},\\
\delta\tilde{\bx} &\triangleq
[\,\delta\bx_b^\top,\ \delta\bx_1^\top,\ \ldots,\ \delta\bx_N^\top\,]^\top\in\mathbb{R}^{4(N+1)}.
\end{align*}
The same stacking convention applies to all other tilded quantities in~\eqref{eq:multi-user_obs}.

Note that the stacked residual vector
$[\,\Delta\tilde{\boldsymbol{\rho}}^\top, \Delta\tilde{\boldsymbol{\Phi}}^\top\,]^\top$
has an associated error vector 
$\tilde{\boldsymbol{\eta}} = [\,\tilde{\boldsymbol{\varepsilon}}^\top,\,\tilde{\boldsymbol{\epsilon}}^\top\,]^\top$
with covariance
\begin{equation}\label{eq:full_cov}
\tilde{\boldsymbol{\Sigma}} = 
\mathbb{E}\{\tilde{\boldsymbol{\eta}}\, \tilde{\boldsymbol{\eta}}^\transpose\} =
\begin{bmatrix}
\tilde{\boldsymbol{\Sigma}}_{\rho} & \tilde{\boldsymbol{\Sigma}}_{\rho,\Phi} \\
\tilde{\boldsymbol{\Sigma}}_{\Phi,\rho} & \tilde{\boldsymbol{\Sigma}}_{\Phi}
\end{bmatrix}
\in \mathbb{R}^{2 K (N+1) \times 2 K (N+1)}.
\end{equation}
Here, $\tilde{\boldsymbol{\Sigma}}_{\rho} = \mathrm{diag}(\boldsymbol{\Sigma}_{\rho,b}, \boldsymbol{\Sigma}_{\rho,1}, \dots, \boldsymbol{\Sigma}_{\rho,N})$,
with $\boldsymbol{\Sigma}_{\rho,r} = \sigma_{\rho,r}^2\mathbf{W}_r^{-1}$ being the pseudorange covariance of receiver~$r$,
$\sigma_{\rho,r}^2$ the known nominal pseudorange-noise variance. An analogous definition applies to $\tilde{\boldsymbol{\Sigma}}_{\Phi}$.
The cross-covariance terms in~\eqref{eq:full_cov} are typically assumed to be negligible, except for baselines above 50 km~\cite[Ch. 2]{medina2022robust}.
Finally, the diagonal weighting matrix $\mathbf{W}_r$ follows standard GNSS variance models based on satellite elevation~\cite{kuusniemi2008user} or signal-to-noise ratio~\cite{medina2018determination}.

%
%

%
%

%
%

\subsubsection{Cooperative DGNSS Model (code-only)}\label{sec:step3}
The full-network SD measurement equations are obtained by applying the SD operator in~\eqref{eq:general_sd_vector} to~\eqref{eq:multi-user_obs} as
\begin{align}\label{eq:single_difference_model}
    \Delta \tilde{\boldsymbol{\rho}}_b &= \tilde{\bD}_b\,\Delta\tilde{\brho} = \tilde{\bH}_N\,\delta\tilde{\bx}_b + \tilde{\boldsymbol{\varepsilon}}_b \in \mathbb{R}^{KN}, \\
    \nonumber \Delta \tilde{\boldsymbol{\Phi}}_b &= \tilde{\bD}_b\,\Delta\tilde{\bPhi} = \tilde{\bH}_N\,\delta\tilde{\bx}_b + \lambda\,\tilde{\bN}_b + \tilde{\boldsymbol{\epsilon}}_b \in \mathbb{R}^{KN},
\end{align}
where common-mode errors, including satellite clock offsets and atmospheric delays, are largely canceled by the single-differencing operation.
%

The covariance of $\tilde{\boldsymbol{\eta}}_b = [\,\tilde{\boldsymbol{\varepsilon}}_b^\top,\ \tilde{\boldsymbol{\epsilon}}_b^\top\,]^\top$ is given by
\begin{equation}\label{eq:sd_cov}
\begin{split}
\tilde{\boldsymbol{\Sigma}}_{b} &= 
\mathbb{E}\{\tilde{\boldsymbol{\eta}}_b\, \tilde{\boldsymbol{\eta}}_b^\transpose\} =
\begin{bmatrix}
\tilde{\boldsymbol{\Sigma}}_{b,\rho} & \tilde{\boldsymbol{\Sigma}}_{b,\rho,\Phi} \\
\tilde{\boldsymbol{\Sigma}}_{b,\Phi,\rho} & \tilde{\boldsymbol{\Sigma}}_{b,\Phi}
\end{bmatrix}
\in \mathbb{R}^{2 K N \times 2 K N},
\end{split}
\end{equation}
and the covariance block associated with pseudorange measurements can be expressed as
\begin{equation}\label{eq:cdgnss_cov}
\tilde{\boldsymbol{\Sigma}}_{b,\rho}
= \tilde{\bD}_b\, \tilde{\boldsymbol{\Sigma}}_\rho\, \tilde{\bD}_b^\top
=
\begin{bmatrix}
\bSigma_{\rho,1}+\bSigma_{\rho,b} & \cdots & \bSigma_{\rho,b} \\
\vdots & \ddots & \vdots \\
\bSigma_{\rho,b} & \cdots & \bSigma_{\rho,N}+\bSigma_{\rho,b}
\end{bmatrix},
\end{equation}
where all off-diagonal terms correspond to $\bSigma_{\rho,b}$
An analogous expression holds for $\tilde{\boldsymbol{\Sigma}}_{b,\Phi}$.
More details regarding the observation model in~\eqref{eq:single_difference_model} can be found in the Supplemental Material, Sec.~S1-C.

%

%
%
\begin{definition}[C-DGNSS Model]\label{def:definition_1}
Having introduced all relevant terms, we now present the complete \gls{C-DGNSS} observation model (see Table~\ref{tab:crb_models}):

\begin{equation}\label{eq:cdgnss_model}
\begin{split}
\Delta\tilde{\brho}_b|\delta\tilde{\mathbf{x}}_b  \sim\mathcal{N}(\tilde{\mathbf{H}}_N\,\delta\tilde{\mathbf{x}}_b , \tilde{\bSigma}_{b,\rho}),\ \delta\tilde{\mathbf{x}}_b \in\mathbb{R}^{4N}.
\end{split}
\end{equation}
%
%
%
%
Here, $\delta\mathbf{x}_{bi}=\delta\mathbf{x}_i-\delta\mathbf{x}_b \overset{(a)}{=} \delta\mathbf{x}_i$, where (a) uses $\mathbf{x}_b=\mathbf{x}_{b,0}$ (surveyed base) $\Rightarrow \delta\mathbf{x}_b=\mathbf{0}$.
Hence, in \gls{C-DGNSS}, the relative state increment reduces to the user’s absolute one as 
$\delta\bx_{bi} = [\,\delta\mathbf{p}_{bi}^\top,\ c\,\delta t_{bi}\,]^\top 
\stackrel{(a)}{=} [\,\delta\mathbf{p}_i^\top,\ c\,\delta t_i\,]^\top$.
\end{definition}

%
%
%
%
%
%

%

%
\subsubsection{Cooperative RTK Model (code and phase)}\label{sec:step4}
%
%
%

%
%
%
Applying the DD operator in~\eqref{eq:operator_dd} to~\eqref{eq:single_difference_model} gives the full-network \gls{DD} measurement equations as 
\begin{align}\label{eq:double_difference_model}
   \Delta \tilde{\boldsymbol{\rho}}_{b}^{p} &= \tilde{\bD}_p\Delta \tilde{\boldsymbol{\rho}}_b = \tilde{\bD}_p\tilde{\mathbf{E}}_N\tilde{\bb} + \tilde{\boldsymbol{\varepsilon}}_b^p \in \mathbb{R}^{(K-1)N} \\
   \nonumber \Delta \tilde{\boldsymbol{\Phi}}_{b}^{p} &= \tilde{\bD}_p\Delta \tilde{\boldsymbol{\Phi}}_b = \tilde{\bD}_p\tilde{\mathbf{E}}_N\tilde{\bb} + \lambda\tilde{\ba} + \tilde{\boldsymbol{\epsilon}}_b^p \in \mathbb{R}^{(K-1)N}
\end{align}
where, beyond the single-difference error cancelation in~\eqref{eq:single_difference_model}, double differencing cancels receiver clock offsets.
Here, $\tilde{\bE}_{N} = \mathbf{I}_{N} \otimes \bE$, and we use the shorthands $\tilde{\bb} = \delta\tilde{\bp}_b$ and $\tilde{\ba} = \tilde{\bN}_b^p$ to denote the network-wide baseline vector and the vector of integer ambiguities, respectively.
Note that, following the DD operator in~\eqref{eq:operator_dd}, we have
\begin{align}\label{eq:worked_ambiguities}
    \tilde{\ba} &= \tilde{\bN}_b^p = [\,\bN_{b1}^{p\top},\ \dots,\ \bN_{bN}^{p\top}\,]^\top \in\mathbb{R}^{(K{-}1)N},\\
    \nonumber \bN_{br}^{p\top} &= [\,N_{br}^{p1},\ \dots,\ N_{br}^{pK}\,]^\top \in\mathbb{R}^{K{-}1},\forall r \in \{1,\dots,N\},\\
    \nonumber N_{br}^{ps} &= N_r^s - N_r^p - (N_b^s - N_b^p)\in\mathbb{R},\forall s \in \{1,\dots,K{-}1\},
\end{align}
and analogous expressions hold for $\Delta\tilde{\boldsymbol{\rho}}_{b}^{p}$ and $\Delta\tilde{\boldsymbol{\Phi}}_{b}^{p}$.

We write the covariance of the stacked error vector $\tilde{\boldsymbol{\eta}}_b^p = [\,\tilde{\boldsymbol{\varepsilon}}_b^{p\top},\ \tilde{\boldsymbol{\epsilon}}_b^{p\top}\,]^\top$ as
\begin{equation}\label{eq:cov_crtk}
\begin{split}
\tilde{\bSigma}_{b}^{p} &=
\begin{bmatrix}
 \tilde{\bSigma}_{b,\rho}^{p} & \tilde{\bSigma}_{b,\rho,\Phi}^{p} \\
 \tilde{\bSigma}_{b,\Phi,\rho}^{p} & \tilde{\bSigma}_{b,\Phi}^{p}
\end{bmatrix} \in \mathbb{R}^{2 (K{-}1) N \times 2 (K{-}1) N},
\end{split}
\end{equation}
where $\tilde{\bSigma}_{b,\rho}^{p} = 
 \tilde{\bD}_p\,\tilde{\boldsymbol{\Sigma}}_{b,\rho}\,\tilde{\bD}_p^\transpose$ and $\tilde{\bSigma}_{b,\Phi}^{p} = 
 \tilde{\bD}_p\,\tilde{\boldsymbol{\Sigma}}_{b,\Phi}\,\tilde{\bD}_p^\transpose$.
More details regarding the observation model in~\eqref{eq:double_difference_model} can be found in the Supplemental Material, Sec.~S1-D.
%
%
%
\begin{definition}[C-RTK Model]\label{def:definition_2}
The complete C-RTK observation model (see Table~\ref{tab:crb_models}) is given by:
%
\begin{equation}\label{eq:crtk_general_model}
\hspace{-0.7cm}
    \begin{split}
     \begin{bmatrix}
    \Delta \tilde{\boldsymbol{\rho}}_{b}^{p} \\
    \Delta \tilde{\boldsymbol{\Phi}}_{b}^{p}
    \end{bmatrix}
    \Bigg|
    \begin{bmatrix}
    \tilde{\bb} \\
    \tilde{\ba}
    \end{bmatrix}
    \sim 
    \mathcal{N}\!\left(\tilde{\bB}\tilde{\bb} + \tilde{\bA}\tilde{\ba},\, \tilde{\bSigma}_{b}^{p}\right),
    \tilde{\bb}\in\mathbb{R}^{3N}, 
    \tilde{\ba}\in\mathbb{R}^{N(K-1)},
    \end{split}
\end{equation}

%
%
%
%
%
\begin{equation}\label{eq:c-rtk_model_mats}
\tilde{\mathbf{A}} = 
\begin{bmatrix}
\tilde{\mathbf{A}}_\rho  \\
\tilde{\mathbf{A}}_\Phi
\end{bmatrix} =
\begin{bmatrix}
\mathbf{0}_{N(K-1)} \\
\lambda \cdot \mathbf{I}_{N(K-1)}
\end{bmatrix},
\tilde{\mathbf{B}} = 
\begin{bmatrix}
\tilde{\mathbf{B}}_\rho  \\
\tilde{\mathbf{B}}_\Phi
\end{bmatrix} =
\begin{bmatrix}
\tilde{\bD}_p\tilde{\bE}_N \\
\tilde{\bD}_p\tilde{\bE}_N
\end{bmatrix},
\end{equation}
with $\tilde{\bD}_p\tilde{\bE}_N = \mathbf{I}_N\otimes\bD\bE$ due to shared satellite geometry among receivers.
%
%
%
Under assumption~(a) in Definition~\ref{def:definition_1} (surveyed base), it follows that \gls{C-RTK} jointly estimates user positions, 
i.e., $\tilde{\bb} = \delta\tilde{\bp}_b \stackrel{(a)}{=} [\,\delta\bp_1^\top,\ \dots,\ \delta\bp_N^\top\,]^\top$, 
and, from~\eqref{eq:worked_ambiguities}, the double-differenced ambiguities with respect to the pivot satellite.


%
%


%



\end{definition}

\begin{remark}[C-DGNSS/C-RTK Model Covariance Structure]\label{remark:3}
We hereafter obtain simplified covariance expressions for the \gls{C-DGNSS} and \gls{C-RTK} models, which underpin the theoretical derivations in Section~\ref{sec:bounds}.

Let us assume that all receivers share identical error characteristics, i.e., \(\sigma_{\rho,r}^2 = \sigma_\rho^2\), \(\sigma_{\Phi,r}^2 = \sigma_\Phi^2\), and \(\mathbf{W}_r = \mathbf{W}\) \(\forall r \in \{1,\ldots,N\}\), leading to \(\boldsymbol{\Sigma}_\rho = \sigma_\rho^2\, \mathbf{W}^{-1}\) and \(\boldsymbol{\Sigma}_\Phi = \sigma_\Phi^2\, \mathbf{W}^{-1}\) for all receivers. Let the base station share the same weighting matrix, i.e., \(\mathbf{W}_b = \mathbf{W}\), but have a different error variance: \(\sigma_{\rho,b}^2 \neq \sigma_\rho^2\), \(\sigma_{\Phi,b}^2 \neq \sigma_\Phi^2\). Under these assumptions, the covariance matrix in~\eqref{eq:cdgnss_cov} simplifies to
\begin{equation}\label{eq:cov_cdgnss_remark}
\tilde{\boldsymbol{\Sigma}}_{b,\rho}^\alpha = 
\begin{bmatrix}
(1+\alpha)\boldsymbol{\Sigma}_\rho & \cdots & \alpha\boldsymbol{\Sigma}_\rho \\
\vdots & \ddots & \vdots \\
\alpha\boldsymbol{\Sigma}_\rho & \cdots & (1+\alpha)\boldsymbol{\Sigma}_\rho
\end{bmatrix}
= \mathbf{C}_N^\alpha \otimes \boldsymbol{\Sigma}_\rho,
\end{equation}
\begin{equation}\label{eq:collab_matrix_alpha}
    \mathbf{C}_N^\alpha = \mathbf{I}_N + \alpha \mathbf{1}_{N},
\end{equation}
being \(\alpha = \sigma_{\rho,b}^2 / \sigma_\rho^2\) the ratio of observation error variances between the base station and the receivers in the network.
An analogous expression holds for $\tilde{\boldsymbol{\Sigma}}_{b,\Phi}^\alpha$.
Assuming uncorrelated code and carrier-phase measurements,~\eqref{eq:sd_cov} simplifies to
\begin{equation}
\tilde{\boldsymbol{\Sigma}}_b^\alpha = 
\begin{bmatrix}
\mathbf{C}_N^\alpha \otimes \boldsymbol{\Sigma}_\rho & \mathbf{0}_{KN} \\
\mathbf{0}_{KN} & \mathbf{C}_N^\alpha \otimes \boldsymbol{\Sigma}_\Phi
\end{bmatrix} \in \mathbb{R}^{2 K N \times 2 K N}.
\end{equation}
and~\eqref{eq:cov_crtk} reduces to
\begin{equation}\label{eq:cov_crtk_remark_1}
\tilde{\boldsymbol{\Sigma}}_b^{p,\alpha} = 
\begin{bmatrix}
\tilde{\boldsymbol{\Sigma}}_{b,\rho}^{p,\alpha} & \mathbf{0}_{(K-1)N} \\
\mathbf{0}_{(K-1)N} & \tilde{\boldsymbol{\Sigma}}_{b,\Phi}^{p,\alpha}
\end{bmatrix} \in \mathbb{R}^{2 (K{-}1) N \times 2 (K{-}1) N},
\end{equation}
\begin{equation}
    \tilde{\boldsymbol{\Sigma}}_{b,\rho}^{p,\alpha} = \mathbf{C}_N^\alpha \otimes \mathbf{D}_p \boldsymbol{\Sigma}_\rho \mathbf{D}_p^\top,
\end{equation}
with an analogous expression for $\tilde{\boldsymbol{\Sigma}}_{b,\Phi}^{p,\alpha}$.
\end{remark}

Note that for \(N = 1\), the cooperative model reduces to the classical observation model for conventional \gls{RTK} positioning:
\begin{equation}
\begin{bmatrix}
\Delta \boldsymbol{\rho}_{b}^{p} \\
\Delta \boldsymbol{\Phi}_{b}^{p}
\end{bmatrix}
\Bigg|
\begin{bmatrix}
\bb \\
\ba
\end{bmatrix}
\sim
\mathcal{N}\!\left(
\mathbf{B}\bb + \mathbf{A}\ba,\,
\bSigma_{b}^{p}
\right),
\quad
\bb \in \mathbb{R}^{n_x},\ 
\ba \in \mathbb{R}^{K-1},
\end{equation}
where \(\bb = \delta\mathbf{p}_b\) denotes the baseline between the user and the reference station, and \(\ba = \bN_b^p\) is the vector of double-differenced integer ambiguities.
The model matrices simplify to
\begin{equation}
\mathbf{A} =
\begin{bmatrix}
\mathbf{A}_\rho \\
\mathbf{A}_\Phi
\end{bmatrix}
=
\begin{bmatrix}
\mathbf{0}_{K-1} \\
\lambda \mathbf{I}_{K-1}
\end{bmatrix},
\quad
\mathbf{B} =
\begin{bmatrix}
\mathbf{B}_\rho \\
\mathbf{B}_\Phi
\end{bmatrix}
=
\begin{bmatrix}
\mathbf{D}_p \mathbf{E} \\
\mathbf{D}_p \mathbf{E}
\end{bmatrix}.
\end{equation}
Similarly, the standard DGNSS model is recovered by considering only \(\Delta \boldsymbol{\rho}_{b}^{p}\).




%

%
\begin{table}[t]
\centering
\caption{Unified representation of parameters of the estimation performance bound for each cooperative GNSS model, mapped to the general FIM expression in~\eqref{eq:unified_model}.}
\label{tab:crb_model_summary}
\renewcommand{\arraystretch}{1.3}
\begin{tabular}{@{}llll@{}}
\toprule
\textbf{Method} & \textbf{Model} & \(\tilde{\bG}\) & \(\tilde{\bR}\) \\
\midrule
C-DGNSS (code-only) & \eqref{eq:bound_cdgnss} & \(\tilde{\bH}_N\) & \(\tilde{\boldsymbol{\Sigma}}_{b,\rho}^\alpha\) \\
C-RTK float & \eqref{eq:fim_crtk_float} & \(\tilde{\bD}_p \tilde{\bE}_N\) & \(\tilde{\boldsymbol{\Sigma}}_{b,\rho}^{p,\alpha}\) \\
C-RTK fix   & \eqref{eq:fim_crtk_fix} & \(\tilde{\bD}_p \tilde{\bE}_N\) & \(\tilde{\boldsymbol{\Sigma}}_{b,\Phi}^{p,\alpha}\) \\
\bottomrule
\vspace{-.9cm}
\end{tabular}
\end{table}

\vspace{-.3cm}
\subsection{Estimators and Performance Bounds}\label{sec:estimation_problem_bounds}
This section presents the \gls{C-DGNSS} (Sec.~\ref{sec:method}-\ref{sec:estimation_problem_bounds}.\ref{sec:est_cdgnss}) and \gls{C-RTK} (Sec.~\ref{sec:method}-\ref{sec:estimation_problem_bounds}.\ref{sec:est_crtk}) estimators, along with their corresponding \gls{CRB}, which provides a theoretical lower bound on estimation variance.
%
%

It is convenient to express the \gls{FIM} of the two models under study in a unified form as
\begin{equation}\label{eq:unified_model}
\bJ(\bomega) = \tilde{\bG}^\transpose \tilde{\bR}^{-1} \tilde{\bG},
\end{equation}
where the model-specific definitions of \(\tilde{\bG}\) and \(\tilde{\bR}\) are summarized in Table~\ref{tab:crb_model_summary}.
%


\begin{remark}[Solvability condition]\label{rmk:solvability_condition}
A sufficient condition for the least squares solution is that the total number of independent measurements must be greater than or equal to the number of unknowns in the system. For the \gls{C-DGNSS} and \gls{C-RTK} estimators, this corresponds to~\cite{7112186} $\sum_{r=1}^{N} K_r\geq 4N$,
where $K_r \leq K$ denotes the number of satellites visible to user $r$. This condition is guaranteed to be satisfied if every user in the scenario has visibility to at least 4 satellites.
\end{remark}

\begin{figure}[t]
\centering

\begin{subfigure}{0.499\textwidth}
    \centering
    \includegraphics[width=.93\textwidth]{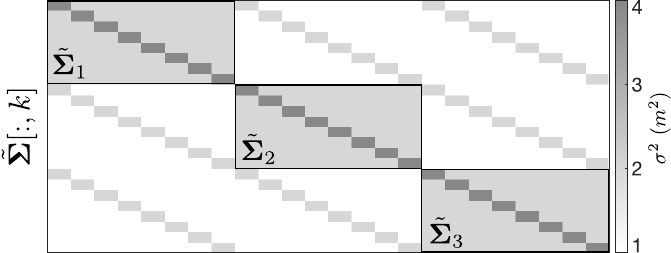}
    \caption{All receivers share the same $K=9$ satellites in view. Matrix subscripts denote user indices.}
    \label{fig:crtk_cov_full}
\end{subfigure}

\vspace{-4pt} 

\begin{subfigure}{0.499\textwidth}
    \centering
    \includegraphics[width=.93\textwidth]{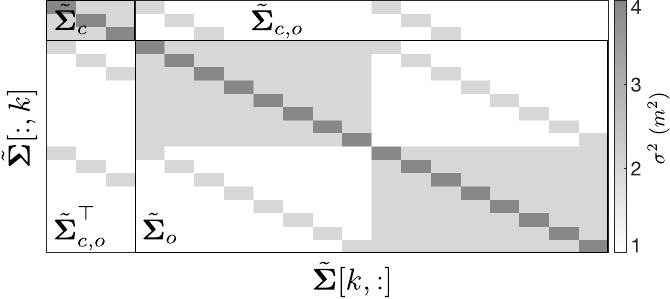}
    \caption{One receiver observes fewer satellites than the rest ($K_\text{c} = 4$); notation for constrained and open-sky users follows Sec.~\ref{sec:bounds}-\ref{sec:network_clustering}.}
    \label{fig:crtk_cov_constrained}
\end{subfigure}

\caption{Heatmaps of the \gls{C-RTK} covariance matrix for pseudorange measurements, $\tilde{\boldsymbol{\Sigma}}_{b,\rho}^{p,\alpha}$, under the assumptions of Remark~\ref{remark:3} (cf.~\eqref{eq:cov_crtk_remark_1}), with \(\bW = \mathrm{I}\), \(\sigma_\rho = 1\) m, $N=3$, and \(\alpha = 1\).
An analogous pattern is obtained for $\tilde{\boldsymbol{\Sigma}}_{b,\Phi}^{p,\alpha}$.
\([k,:]\) and \([:,k]\) denote row and column slices.
See~\cite[Fig.~3]{calatrava_cdgnss_2025} for the \gls{C-DGNSS} case under $\alpha=1$. See Supplemental Material (Sec.~S1-D) for a detailed explanation of the element-wise noise contributions in C-RTK.}
\label{fig:covariance_motivation}

\vspace{-.6cm}
\end{figure}
\vspace{-.3cm}

\subsubsection{Cooperative DGNSS Estimator and Bound}\label{sec:est_cdgnss}
Following the code-only observation model in~\eqref{eq:cdgnss_model}, and if the condition in Remark~\ref{rmk:solvability_condition} is met, \gls{C-DGNSS} can be formulated as a \gls{WLS} problem:
\begin{equation}\label{eq:cdgnss_estimator}
\hat{\bomega} = \mathop{\mathrm{arg\,min}}_{\hat{\bomega} \in \mathbb{R}^{M_y}} \left\{(\by - \tilde{\bG}\bomega)^\transpose\tilde{\bR}^{-1}(\by - \tilde{\bG}\bomega)\right\}.
\end{equation}
Recall that the terms in this formulation are detailed for the \gls{C-DGNSS} model in Tables~\ref{tab:crb_models} and~\ref{tab:crb_model_summary}.

The estimator in~\eqref{eq:cdgnss_estimator} can be obtained through an iterative procedure that relinearizes the system model at each step, equivalent to the Gauss–Newton optimization method~\cite{bell1993iterated}.
%
%
%
%
%
At the \(k\)-th iteration, one has
\begin{equation}\label{eq:cdgnss_estimator_update}
\hat{\bomega}^{k+1} = \hat{\bomega}^k 
+ \left(\tilde{\bG}^{k^\top}\tilde{\bR}^{-1}\tilde{\bG}^k\right)^{-1}
\tilde{\bG}^{k^\top}\tilde{\bR}^{-1}(\by-\tilde{\bG}^k\hat{\bomega}^k),
\end{equation}
where the geometry matrix \(\tilde{\bG}^k\) is evaluated at \(\hat{\bomega}^k\), although this dependency is omitted in the notation for simplicity.
%
Remarkably, the computational cost of the proposed estimator in~\eqref{eq:cdgnss_estimator_update} is dominated by the inversion of $(\tilde{\mathbf{G}}^{\top}\tilde{\mathbf{R}}^{-1}\tilde{\mathbf{G}})$, 
which scales as $\mathcal{O}(N^3)$, increasing rapidly with the number of users and motivating future work on distributed formulations to improve scalability.

Given the Gaussian assumption in~\eqref{eq:cdgnss_model}, the estimator in~\eqref{eq:cdgnss_estimator} is optimal and equivalent to the \gls{MLE}, whose estimation accuracy is theoretically lower-bounded by the \gls{CRB}.
To evaluate the CRB for a specific user \(r\), we extract the corresponding \(3 \times 3\) submatrix from the \gls{FIM}, which is associated with that user's 3D position parameters. 
For this purpose, we introduce the operator~\cite{calatrava_cdgnss_2025,calatrava2025towards}
\begin{equation}\label{eq:operator}
    \blockj{r}(\mathbf{X}) = \mathbf{X}{[3(r-1)+1 : 3r,\ 3(r-1)+1 : 3r]},
\end{equation}
where \([i\!:\!j,\ k\!:\!l]\) denotes the submatrix formed by rows \(i\) to \(j\) and columns \(k\) to \(l\), and the indices follow a one-based convention, with \(r \in \{1, 2, \dots, N\}\).
A lower bound on the variance of the estimate of the \(i\)-th parameter is provided by the CRB as
\begin{equation}
\text{Var}(\textcolor{black}{\hat\bomega}_i) \geq \mathbf{CRB}(\textcolor{black}{\bomega}_i) = [\bJ^{-1}(\bomega)]_{ii},
%
\end{equation}
where $\bomega_i=[\bomega]_i$, and the \gls{FIM} is defined as 
\begin{equation}\label{eq:bound_cdgnss}
\begin{split}
[\bJ(\bomega)]_{i,j} = -\mathbb{E}\left[\frac{\partial^2\ln p(\by;\bomega)}{\partial \textcolor{black}{\bomega}_i\partial\bomega_j}\right] = [\tilde{\bG}^\transpose \tilde{\bR}^{-1} \tilde{\bG}]_{i,j}.
\end{split}
\end{equation}
As a scalar 3D positioning performance metric for receiver~$r$ we calculate the \gls{RMSE} of its position estimate, expressed in meters, as $\sqrt{\mathrm{Tr}\left[\blockj{r}\left(\bJ^{-1}(\bomega)\right)\right]}
$.

%

\vspace{-.15cm}

\subsubsection{Cooperative RTK Estimator and Bound}\label{sec:est_crtk}



The model in~\eqref{eq:unified_model_1} defines a mixed linear regression problem involving both real-valued parameters $\boldsymbol{\omega}$ and integer-valued parameters $\boldsymbol{z}$, a characteristic feature of GNSS carrier-phase positioning~\cite{medina2021cramer}. Due to the integer nature of $\boldsymbol{z}$, the estimation problem admits no closed-form solution. Following standard RTK practice, a three-stage procedure is adopted (i.e., float estimation, integer ambiguity resolution, and fixed estimation) and this architecture is extended to the cooperative setting in~\cite{medina2023collaborative}.

In~\cite{medina2021cramer}, the performance of this mixed real–integer estimation problem is characterized by a \gls{MSB} constructed from a finite set of test points. That framework provides a rigorous treatment of both float and fixed regimes but leads to expressions that are not easily embedded in the unified C-DGNSS/C-RTK analysis introduced in this work. We therefore focus on the corresponding Cramér–Rao-type bounds that describe the float and fix regions of C-RTK, which can be interpreted as limiting cases of the MSB when the ambiguities remain unresolved (float) or are correctly fixed. For completeness, we briefly summarize these bounds below and refer the reader to~\cite{medina2021cramer} for the full MSB formulation.

Note that in addition to the CRB, the probability of success of integer ambiguity resolution (i.e., the likelihood that the correct integer vector is fixed) can be upper bounded as in~\cite[Eq.~(17)]{medina2023collaborative}.

\textbf{\textbf{C-RTK Float:}} When the integer ambiguities are treated as real-valued, i.e., \(\bomega \in \mathbb{R}^{K_\omega}\), \(\bz \in \mathbb{R}^{K_z}\), C-RTK operates in the float regime. In this case, the FIM for $\boldsymbol{\theta}$ is given by (see~\cite[Eq.~(26)]{medina2021cramer} for the non-cooperative case):
%
%
%
%
\begin{equation}\label{eq:fim_crtk_float}
\begin{split}
&\bJ(\btheta) =
\begin{bmatrix}
\tilde{\mathbf{B}}^{\top} \\
\tilde{\mathbf{A}}^{\top}
\end{bmatrix}
(\tilde{\bSigma}_{b}^{p,\alpha})^{-1}
\begin{bmatrix}
\tilde{\mathbf{B}} & \tilde{\mathbf{A}}
\end{bmatrix}
\\&=
\begin{bmatrix}
\tilde{\mathbf{B}}_{\rho}^{\top} & \tilde{\mathbf{B}}_{\Phi}^{\top} \\
\tilde{\mathbf{A}}_{\rho}^{\top} & \tilde{\mathbf{A}}_{\Phi}^{\top}
\end{bmatrix}
\begin{bmatrix}
(\tilde{\bSigma}_{b,\rho}^{p,\alpha})^{-1} & \\
& (\tilde{\bSigma}_{b,\Phi}^{p,\alpha})^{-1}
\end{bmatrix}
\begin{bmatrix}
\tilde{\mathbf{B}}_{\rho} & \tilde{\mathbf{A}}_{\rho} \\
\tilde{\mathbf{B}}_{\Phi} & \tilde{\mathbf{A}}_{\Phi}
\end{bmatrix}
\\ &=
\begin{bmatrix}
\tilde{\mathbf{B}}_{\rho}^{\top} \left( (\tilde{\boldsymbol{\Sigma}}_{b,\rho}^{p,\alpha})^{-1}+(\tilde{\boldsymbol{\Sigma}}_{b,\Phi}^{p,\alpha})^{-1} \right) \tilde{\mathbf{B}}_{\rho} &
\lambda \tilde{\mathbf{B}}_{\rho}^{\top} (\tilde{\boldsymbol{\Sigma}}_{b,\Phi}^{p,\alpha})^{-1} \\
\lambda (\tilde{\boldsymbol{\Sigma}}_{b,\Phi}^{p,\alpha})^{-1} \tilde{\mathbf{B}}_{\rho} &
\lambda^{2} (\tilde{\boldsymbol{\Sigma}}_{b,\Phi}^{p,\alpha})^{-1}
\end{bmatrix}.
\end{split}
\end{equation}
Here, we incorporate $\alpha$ as defined in Remark~\ref{remark:3}, assume uncorrelated carrier-phase and code measurements following~\eqref{eq:cov_crtk_remark_1}, and adopt from~\eqref{eq:c-rtk_model_mats} $\tilde{\mathbf{A}}_{\Phi} = \lambda\,\mathbf{I}_{N(K-1)}$, $\tilde{\mathbf{A}}_{\rho}=\mathbf{0}$, and $\tilde{\mathbf{B}}_{\rho}=\tilde{\mathbf{B}}_{\Phi}$.
Let $\mathbf{U}$, $\mathbf{V}$, and $\mathbf{W}$ denote the top-left, off-diagonal, and bottom-right blocks of~\eqref{eq:fim_crtk_float}, respectively. The CRB for $\boldsymbol{\omega}$ corresponds to the $(1,1)$ block of $\mathbf{J}(\boldsymbol{\theta})^{-1}$, which by the Schur complement is
\begin{equation}\label{eq:crb_float}
\begin{split}
    \mathbf{CRB}(\hat\bomega)
     = \big(\mathbf{U} - \mathbf{V} \mathbf{W}^{-1} \mathbf{V}^{\top}\big)^{-1} = \left( \tilde{\mathbf{B}}_{\rho}^{\top}
    \big(\tilde{\boldsymbol{\Sigma}}_{b,\rho}^{p,\alpha}\big)^{-1}
    \tilde{\mathbf{B}}_{\rho} \right)^{-1}.
\end{split}
\end{equation}
In the float regime, precision is dominated by the code measurements because the ambiguities remain unresolved. The RTK float bound corresponds to the information obtained after projecting the observation matrix onto the subspace orthogonal to the ambiguity space~\cite[Eq. (2.16)]{medina2022robust}.

%

%
\textbf{\textbf{C-RTK Fix:}} Further gains arise once the ambiguities are correctly resolved and treated as integers. In this fix region, the FIM for $\boldsymbol{\omega}$ is given by~\cite[Eq.~(25a)]{medina2021cramer}

\begin{equation}\label{eq:fim_crtk_fix}
\begin{split}
    \bJ(\bomega) &= 
\tilde{\mathbf{B}}^{\top}
(\tilde{\bSigma}_b^{p,\alpha})^{-1} \tilde{\mathbf{B}} =  \tilde{\mathbf{B}}_{\rho}^{\top} \left( (\tilde{\boldsymbol{\Sigma}}_{b,\rho}^{p,\alpha})^{-1}+(\tilde{\boldsymbol{\Sigma}}_{b,\Phi}^{p,\alpha})^{-1} \right)\tilde{\mathbf{B}}_{\rho},
%
\end{split}
\end{equation}
where the assumptions and matrix definitions from~\eqref{eq:fim_crtk_float} continue to apply.
Under the standard assumption that $\sigma_\Phi \ll \sigma_\rho$, the approximation
$(\tilde{\boldsymbol{\Sigma}}_{b,\Phi}^{p,\alpha})^{-1} + (\tilde{\boldsymbol{\Sigma}}_{b,\rho}^{p,\alpha})^{-1} \approx (\tilde{\boldsymbol{\Sigma}}_{b,\Phi}^{p,\alpha})^{-1}$
holds, leading to an estimation precision dominated by the carrier-phase measurements, with the bound
\begin{equation}\label{eq:crb_freq}
    \mathbf{CRB}(\hat\bomega) \approx  \left(\tilde{\mathbf{B}}_{\rho}^{\top}(\tilde{\boldsymbol{\Sigma}}_{b,\Phi}^{p,\alpha})^{-1}\tilde{\mathbf{B}}_{\rho}\right)^{-1}.
\end{equation}
%
%
Thus far in this work, we have introduced unified observation models for cooperative DGNSS and cooperative RTK, along with their position estimation problems and CRB-based performance bounds.

\section{Theoretical Performance Analysis}\label{sec:bounds}
This section provides a comprehensive theoretical analysis of
the asymptotic performance of the C-DGNSS and C-RTK estimators as a function of three key parameters relevant to the cooperative positioning scenario considered in this work, as illustrated in Fig.~\ref{fig:intro_cdgnss}: \textit{(i)} the number of aiding users in the network \(N_\text{o}\); \textit{(ii)} the number of satellites exclusively visible to them \(K_\text{o}\); and \textit{(iii)} the variance ratio \(\alpha\), introduced in Remark~\ref{remark:3}.

Sec.~\ref{sec:bounds}-\ref{sec:network_clustering} introduces the two-cluster visibility model considered in this work.
In Sec.~\ref{sec:bounds}-\ref{sec:bounds:fisher}, we provide the closed-form expression of the \gls{FIM} in~\eqref{eq:unified_model} parameterized as a function of \(N_\text{o}\), \(K_\text{o}\), and $\alpha$, with the detailed steps given in Sec.~S2 of the Supplemental Material. Sec.~\ref{sec:bounds}--\ref{sec:bounds:analysis} presents the theoretical asymptotic analysis.

\vspace{-.3cm}
\subsection{Network Clustering}\label{sec:network_clustering}
We address the inherent heterogeneity of satellite visibility in practical GNSS deployments by classifying the cooperative network into two clusters. 

The first cluster comprises \(N_\text{c}\) \textit{constrained-visibility users} (also referred to as \textit{aided users}), operating under constrained visibility conditions, such as in urban environments, with access to \(K_\text{c}\) satellites. The second cluster consists of the remaining \(N_\text{o} = N - N_\text{c}\) \textit{open-sky users} (also referred to as \textit{aiding users}), who observe a total of \(K = K_\text{c} + K_\text{o}\) satellites, where \(K_\text{o} \geq 0\) denotes the number of satellites exclusively visible to this group.
%
%

Figure~\ref{fig:covariance_motivation} depicts how the covariance structure formulated in Definitions~\ref{def:definition_1} and~\ref{def:definition_2} varies between a homogeneous and a clustered network, the latter including users with constrained visibility.

To incorporate the clustered geometry into the unified cooperative model in~\eqref{eq:unified_model_1}, we define
\begin{equation}\label{eq:obs_aiding}
    \tilde{\bG} = \operatorname{blkdiag}(\mathbf{G}_c,\, \mathbf{I}_{N_\text{o}} \otimes \mathbf{G}),\ \bG = [\bG_c^\top\ \bG_o^\top]^\top,
\end{equation}
%
%
\begin{equation}\label{eq:sigma_clusters}
\begin{split}
\tilde{\bR} &= 
%
\begin{bmatrix}
(1+\alpha)\mathbf{R}_{c}
&
\underline{1}_{N_{o}}^\top \otimes 
\begin{bmatrix}
\alpha\mathbf{R}_{c} & \mathbf{0}_{K_{c},\,K_{o}}
\end{bmatrix}
\\[1em]
\underline{1}_{N_{o}} \otimes 
\begin{bmatrix}
\alpha\mathbf{R}_{c} \\
\mathbf{0}_{K_{o},\,K_{c}}
\end{bmatrix}
&
\mathbf{C}^\alpha_{N_{o}} \otimes \mathbf{R}
\end{bmatrix}.
\\
%
%
%
%
\end{split}
\end{equation}
where $\operatorname{blkdiag}(\cdot)$ denotes the block-diagonal concatenation operator, \(\bR = \begin{bmatrix}
\bR_\text{c} & \mathbf{0}  \\
\mathbf{0} &  \bR_\text{o}
\end{bmatrix}\), and $\mathbf{C}^\alpha_{N_\text{o}}$ is as introduced in~\eqref{eq:collab_matrix_alpha}.
Let us denote the blocks of matrix $\tilde{\mathbf{R}}$ as 
$\tilde{\mathbf{R}}_{c} = (1+\alpha)\mathbf{R}_{c}$, 
$\tilde{\mathbf{R}}_{\mathrm{c,o}} = \mathbf{1}_{N_{o}}^\top \otimes 
\begin{bmatrix}
\alpha\mathbf{R}_{c} & \mathbf{0}_{K_{c},\,K_{o}}
\end{bmatrix}$, and $\tilde{\mathbf{R}}_{o} = \mathbf{C}^{\alpha}_{N_{o}} \otimes \mathbf{R}$.
Furthermore, in~\eqref{eq:sigma_clusters}, we adopt the assumptions outlined in Remark~\ref{remark:3} and consider the case \(N_\text{c} = 1\), which facilitates the analytical manipulation of the covariance expression while still providing a representative assessment of the position estimation performance for an aided user. This performance is quantified by
\begin{equation}\label{eq:performance_assessment_bound}
    \mathbf{CRB}(\hat{\bomega}_c) = \blockj{1}(\bJ(\bomega)^{-1}),
\end{equation}
where $\bomega_c \in\mathbb{R}^3$ denotes the real-valued unknowns associated with the aided user.
Let us introduce
\begin{equation}\label{eq:fims}
    \bJ_\text{c} = \bG_\text{c}^\top \bR_\text{c}^{-1} \bG_\text{c}, \quad
    \bJ_\text{o} = \bG_\text{o}^\top \bR_\text{o}^{-1} \bG_\text{o},
\end{equation}
where $\bJ_\text{c}$ and $\bJ_\text{o}$ capture the Fisher information from jointly observed satellites and from satellites visible only to the aiding users, respectively.

\textbf{\textbf{Theoretical Benchmarks:}} We consider two benchmarks for comparison. The first is the non-cooperative bound reflecting the performance of standard differential architectures, i.e., \gls{DGNSS}, \gls{RTK}, defined as
\begin{equation}\label{eq:bound:noncoop}
    \mathbf{CRB}_\text{Non-Coop} = (1+\alpha)\,\bJ_\text{c}^{-1}.
\end{equation}
Here, the factor \(1+\alpha\) accounts for the contribution of error variances from both the base station and the aided user, as reflected in the diagonal elements of~\eqref{eq:cov_cdgnss_remark}.
The second is the \textit{ideal bound}, which is defined as
\begin{equation}\label{eq:bound:ideal}
\mathbf{CRB}_\text{Ideal} = \bJ_\text{c}^{-1}.
\end{equation}
This corresponds to a differential architecture with noiseless base station, i.e., $\alpha=0$, yielding to a performance bounded by he geometry of the aided user.

\subsection{Parameterized Fisher Information Matrix}\label{sec:bounds:fisher}

From the unified linear model in~\eqref{eq:unified_model}, the FIM admits the parameterization in~\eqref{eq:inverse_bound} for arbitrary \(N_{\text{o}}\) and \(\alpha\), owing to the block structure of \(\tilde{\mathbf{R}}\).
The intermediate derivation steps leading to~\eqref{eq:inverse_bound} and the coefficients $\beta_i(\alpha)$, $i \in \{0,1,\dots,5\}$ are provided in Sec.~S2 of the Supplemental Material.
The compact and full definitions of $\beta_i(\alpha)$ are listed 
in Table~\ref{tab:beta_coefficients}; $\beta_3(\alpha)$ is only used in the supplement.

Left/right-multiplication by $\tilde{\mathbf{G}}$ preserves the block structure of \(\tilde{\mathbf{R}}\), mapping each covariance block into its corresponding block 
of $\mathbf{J}(\boldsymbol{\omega})$. 
The diagonal blocks scale with $\mathbf{J}_{c}$ and $\mathbf{J}_{o}$, reflecting the contributions 
of common and exclusive satellites, whereas the off-diagonal blocks depend solely on 
$\mathbf{J}_{c}$. 
This results in the compact expression in~\eqref{eq:inverse_bound}, whose sub-blocks 
$\bM_{ij}$ for $i,j\in\{1,2\}$ will be used in the asymptotic analysis of 
Sections~\ref{sec:bounds}--\ref{sec:bounds:analysis}. 
For the compact form of block $\mathbf{M}_{22}$, we use that $\beta_1 = \beta_2 + 1$ 
and $\beta_4 = \beta_5 + 1$, as explained in the supplement.

\begin{figure*}[ht]
\begin{equation}\label{eq:inverse_bound}
\small
\begin{split}
\bJ(\bomega)
&=
\left[
\begin{array}{c@{\hspace{1.5em}}c}
\beta_1(\alpha)\mathbf{J}_\text{c}
&
\begin{bmatrix}
\beta_2(\alpha)\mathbf{J}_\text{c} & \cdots & \beta_2(\alpha)\mathbf{J}_\text{c}
\end{bmatrix}
\\[2.5ex]
\begin{bmatrix}
\beta_2(\alpha)\mathbf{J}_\text{c} \\
\vdots \\
\beta_2(\alpha)\mathbf{J}_\text{c}
\end{bmatrix}
&
\begin{bmatrix}
\beta_{1}(\alpha)\mathbf{J}_\text{c} + \beta_4(\alpha)\mathbf{J}_\text{o} & \cdots & \beta_{2}(\alpha)\mathbf{J}_\text{c} + \beta_5(\alpha)\mathbf{J}_\text{o} \\
\vdots & \ddots & \vdots \\
\beta_{2}(\alpha)\mathbf{J}_\text{c} + \beta_5(\alpha)\mathbf{J}_\text{o} & \cdots & \beta_{1}(\alpha)\mathbf{J}_\text{c} + \beta_4(\alpha)\mathbf{J}_\text{o}
\end{bmatrix}
\end{array}
\right]
=
\left[
\begin{array}{c@{\hspace{1.5em}}c}
\underbrace{\beta_1(\alpha)\mathbf{J}_\text{c}}_{\mathbf{M}_{11}}
&
\underbrace{\beta_2(\alpha)\left(\b1^\transpose_{N_\text{o}}\otimes\mathbf{J}_\text{c}\right)}_{\mathbf{M}_{12}} 
\\[2.5ex]
\underbrace{\beta_2(\alpha)\left(\b1_{N_\text{o}}\otimes\mathbf{J}_\text{c}\right)}_{\mathbf{M}_{21}}
&
\underbrace{%
\begin{aligned}
\left(\bI_{N_\text{o}} + \beta_{2}(\alpha)\,\mathbf{1}_{N_\text{o}}\right)\!\otimes\bJ_\text{c} \\
+\; \left(\bI_{N_\text{o}} + \beta_{5}(\alpha)\,\mathbf{1}_{N_\text{o}}\right)\!\otimes\bJ_\text{o}
\end{aligned}
}_{\mathbf{M}_{22}}
\end{array}
\right]
\\
%
&=
\left[
\begin{array}{c@{\hspace{1.5em}}c}
\displaystyle \frac{\alpha N_\text{o} + 1}{\alpha N_\text{o} + \alpha + 1} \mathbf{J}_\text{c}  &
\begin{bmatrix}
\displaystyle -\frac{\alpha}{\alpha N_\text{o} + \alpha + 1} \mathbf{J}_\text{c} &
\cdots &
\displaystyle -\frac{\alpha}{\alpha N_\text{o} + \alpha + 1} \mathbf{J}_\text{c}
\end{bmatrix}
\\[2ex]
\begin{bmatrix}
\displaystyle -\frac{\alpha}{\alpha N_\text{o} + \alpha + 1} \mathbf{J}_\text{c} \\
\vdots \\
\displaystyle -\frac{\alpha}{\alpha N_\text{o} + \alpha + 1} \mathbf{J}_\text{c}
\end{bmatrix}
&
\begin{bmatrix}
\displaystyle \frac{\alpha N_\text{o} + 1}{\alpha N_\text{o} + \alpha + 1} \mathbf{J}_\text{c} + \frac{1 + \alpha(N_\text{o} - 1)}{\alpha N_\text{o} + 1} \mathbf{J}_\text{o} &
\cdots &
\displaystyle -\frac{\alpha}{\alpha N_\text{o} + \alpha + 1} \mathbf{J}_\text{c} - \frac{\alpha}{\alpha N_\text{o} + 1} \mathbf{J}_\text{o}
\\
\vdots & \ddots & \vdots \\
\displaystyle -\frac{\alpha}{\alpha N_\text{o} + \alpha + 1} \mathbf{J}_\text{c} - \frac{\alpha}{\alpha N_\text{o} + 1} \mathbf{J}_\text{o} &
\cdots &
\displaystyle \frac{\alpha N_\text{o} + 1}{\alpha N_\text{o} + \alpha + 1} \mathbf{J}_\text{c} + \frac{1 + \alpha(N_\text{o} - 1)}{\alpha N_\text{o} + 1} \mathbf{J}_\text{o}
\end{bmatrix}
\end{array}
\right]
\end{split}
\end{equation}
\vspace{-.5cm}
\caption*{The detailed block-matrix and Kronecker-product manipulations are deferred to Sec. S2 of the Supplemental Material.}
\vspace{-.5cm}
\vspace{0.5em}
\noindent\rule{\textwidth}{0.4pt}
\vspace{0.5em}
\vspace{-.8cm}
\end{figure*}

\begin{table}[t]
\centering
\caption{$\beta_i(\alpha)$ coefficient definitions as introduced in Sec.~\ref{sec:bounds}--\ref{sec:bounds:fisher}. Note that $\beta_1 = \beta_2 + 1$ 
and $\beta_4 = \beta_5 + 1$. Derivations leading to these expressions can be found in Sec.~S2 of the Supplemental Material.}
\label{tab:beta_coefficients}
\renewcommand{\arraystretch}{2.2}
\begin{tabular}{@{}c c c@{}}
\toprule
\noalign{\vspace{-2ex}}
\textbf{Coefficient} & \textbf{Compact Form} & \textbf{Full Form} \\
\noalign{\vspace{-6ex}}

\\
\midrule
\noalign{\vspace{-1.7ex}}
$\beta_0(\alpha)$ 
& --- 
& $\displaystyle \frac{1}{1 + \alpha N_o}$ 
\\

$\beta_1(\alpha)$ 
& $\displaystyle \left(\alpha \beta_0(\alpha) + 1\right)^{-1}$
& $\displaystyle \frac{\alpha N_o + 1}{\alpha N_o + \alpha + 1}$ 
\\

$\beta_2(\alpha)$ 
& $\displaystyle -\alpha \,\beta_1(\alpha)\,\beta_0(\alpha)$
& $\displaystyle -\frac{\alpha}{\alpha N_o + \alpha + 1}$ 
\\

$\beta_3(\alpha)$ 
& $\displaystyle \alpha\,\beta_2(\alpha)\,\beta_0(\alpha)$
& $-\frac{\alpha^2}{(1+\alpha N_o)\bigl(\alpha N_o + \alpha + 1\bigr)}$ 
\\

$\beta_4(\alpha)$ 
& $\displaystyle 1 - \alpha\,\beta_0(\alpha)$
& $\displaystyle \frac{1 + \alpha (N_o - 1)}{\alpha N_o + 1}$ 
\\

$\beta_5(\alpha)$ 
& $\displaystyle -\alpha\,\beta_0(\alpha)$
& $\displaystyle -\frac{\alpha}{\alpha N_o + 1}$ 
\\

\bottomrule
\end{tabular}
\end{table}

\vspace{-.35cm}

%

\subsection{Study of Asymptotic Regimes}\label{sec:bounds:analysis}
This section focuses on the term in~\eqref{eq:performance_assessment_bound} capturing the estimation performance of the aided user, which can be calculated with the Schur complement of matrix~\eqref{eq:inverse_bound} as
\begin{equation}\label{eq:bound_main}
\begin{split}
    \mathbf{CRB}(\hat{\bomega}_c)
    = \left(\bM_{11} - \bM_{12} \bM_{22}^{-1} \bM_{21}\right)^{-1}.
\end{split}
\end{equation}
We next examine this bound under key limiting regimes. Note that for $N_\text{o}=0$, the bound reduces to the non-cooperative case and needs no further analysis.


It can be shown that the CRB for code-based DGNSS and RTK float are equivalent in the absence of ephemeris or satellite clock errors. Hence, the asymptotic analysis based on the C-DGNSS block structure in~\eqref{eq:sigma_clusters} also characterizes the asymptotic regimes of C-RTK float. Moreover, since the C-RTK fixed bound in~\eqref{eq:crb_freq} has the same analytical form, differing only by the measurement precision, these conclusions also extend to C-RTK fixed.

\subsubsection{Ideal Aiding-Cluster Satellite Visibility}\label{sec:bounds:analysis:ideal}
The limit $\mathbf{J}_\text{o} \to \infty$ models ideal satellite visibility for the aiding users, i.e., an aiding cluster whose exclusive satellites provide arbitrarily large Fisher information.

%
%
Considering that $\bM_{22}^{-1}\xrightarrow[\mathbf{J}_\text{o} \to \infty]{}\mathbf{0}$\footnote{We write the bottom-right block of~\eqref{eq:inverse_bound} as $\mathbf{M}_{22}(\lambda) = \mathbf{A} + \lambda \mathbf{B}$, where $\mathbf{A}$ and $\mathbf{B}$ contain the $\mathbf{J}_\text{c}$ and $\mathbf{J}_\text{o}$ terms, and $\lambda>0$. 
As $\lambda \to \infty$, the $\lambda\mathbf{B}$ term dominates, yielding $\mathbf{M}_{22}^{-1}(\lambda) \to \tfrac{1}{\lambda}\mathbf{B}^{-1}$.},~\eqref{eq:bound_main} simplifies to
\begin{equation}\label{eq:proof_1}
\begin{split}
    \mathbf{CRB}(\hat{\bomega}_c) \xrightarrow[\mathbf{J}_\text{o} \to \infty]{} \beta_1(\alpha)^{-1} \bJ_\text{c} ^{-1}
\leq \mathbf{CRB}_\text{Non-Coop},
\end{split}
\end{equation}
where the inequality holds for all \(N_\text{o} \in \mathbb{N}_0\) and $\alpha>0$.
This result shows that, under ideal aiding visibility, cooperation always yields an improvement over the non-cooperative bound in~\eqref{eq:bound:noncoop}.
%
%
%

The term $\beta_1(\alpha)^{-1}$ decreases monotonically with \(N_\text{o}\), indicating that the cooperative bound becomes increasingly tighter as more aiding users are added.
This monotonic decrease is confirmed by $\frac{\partial \beta_1(\alpha)^{-1}}{\partial N_\text{o}} 
= -\frac{\alpha^2}{(\alpha N_\text{o} + 1)^2}$ being strictly negative for all \(\alpha > 0\).
For a fixed $N_\text{o}$ and $\alpha_1 > \alpha_2$, we have $\left|\frac{\partial \beta_1(\alpha_1)^{-1}}{\partial N_\text{o}}\right| > \left|\frac{\partial \beta_1(\alpha_2)^{-1}}{\partial N_\text{o}}\right|$, meaning that $\beta_1(\alpha)^{-1}$ decreases faster with $N_\text{o}$ when $\alpha$ is larger. In other words, the rate of improvement due to cooperation is more noticeable for noisier reference stations.



\subsubsection{Homogeneous Inter-Cluster Satellite Visibility}
%
%
We examine the homogeneous case $\mathbf{J}_\text{o}=\mathbf{0}$, where all users have identical visibility and no aiding user provides additional information. This serves as a baseline to test whether cooperation can help when visibility offers no diversity.
The CRB under homogeneous satellite visibility is given by:
\begin{equation}\label{eq:homogeneous_visibility_bound}
    \left. \mathbf{CRB}(\hat{\bomega}_c) \right|_{\mathbf{J}_\text{o} = \mathbf{0}}
= 
\left( \mathbf{M}_{11}
- \bM_{12} \left[\left. \mathbf{M}_{22} \right|_{\mathbf{J}_\text{o} = \mathbf{0}} \right]^{-1} \bM_{21} \right)^{-1}. 
\end{equation}
Let us calculate the intermediate terms as\footnote{In~\eqref{eq:M22_inversion:b} and~\eqref{eq:m12_M22_inv:b}, 
we use the standard Kronecker identities 
$(A\otimes B)^{-1}=A^{-1}\otimes B^{-1}$ and 
$(A\otimes B)(C\otimes D)=(AC)\otimes(BD)$.}
\begin{subequations}\label{eq:M22_inversion}
\begin{align}
\left[\left.\mathbf{M}_{22}\right|_{\mathbf{J}_\text{o} = \mathbf{0}}\right]^{-1}
&=
\left[\left(\bI_{N_\text{o}} + \beta_{2}(\alpha)\,\mathbf{1}_{N_\text{o}}\right)\!\otimes\bJ_\text{c}\right]^{-1}
\label{eq:M22_inversion:a}
\\
&=
\left[\mathbf{I}_{N_\text{o}}
+\beta_{2}(\alpha)\mathbf{1}_{N_\text{o}}\right]^{-1}
\otimes\bJ_\text{c}^{-1}
\label{eq:M22_inversion:b}
\\
&=
\underbrace{\left(
\bI_{N_\text{o}}
-
\frac{\beta_{2}(\alpha)}{\beta_{2}(\alpha)N_\text{o}+1}
\mathbf{1}_{N_\text{o}} 
\right)}_{\bOmega_1}
\otimes\ \bJ_\text{c}^{-1},
\label{eq:M22_inversion:c}
\end{align}
\end{subequations}
\begin{subequations}\label{eq:m12_M22_inv}
\begin{align}
\bM_{12}\left[\left.\mathbf{M}_{22}\right|_{\mathbf{J}_\text{o} = \mathbf{0}}\right]^{-1}
&=
\beta_2(\alpha)\left(\underline{1}^\transpose_{N_\text{o}}\otimes\mathbf{J}_\text{c}\right)
\left(\bOmega_1\otimes\bJ_\text{c}^{-1}\right)
\label{eq:m12_M22_inv:a}
\\
&=
\beta_2(\alpha)\bigl(\underline{1}^\transpose_{N_\text{o}}\bOmega_1\bigr)
    \otimes
\left(\mathbf{J}_\text{c}\bJ_\text{c}^{-1}\right)
\label{eq:m12_M22_inv:b}
\\
&=
-\frac{\alpha}{\alpha+1}\,
\underline{1}_{N_\text{o}}^\transpose \otimes \bI_{K_\text{c}}.
\label{eq:m12_M22_inv:c}
\end{align}
\end{subequations}
The inversion in~\eqref{eq:M22_inversion:c} follows directly from the 
Sherman--Morrison formula:
\begin{equation}\label{eq:sherman_morrison}
    (\mathbf{I} + \mathbf{u}\mathbf{v}^\top)^{-1}
    = \mathbf{I} - 
    \frac{\mathbf{u}\mathbf{v}^\top}{1 + \mathbf{v}^\top \mathbf{u}},
\end{equation}
applied with $\bu=\bv=\sqrt{\beta_{2}(\alpha)}\,\underline{1}_{N_\text{o}}$, and leading to the auxiliary matrix $\bOmega_1$ with diagonal entries $\frac{1+2\alpha}{1+\alpha}$ and 
off-diagonal entries $\frac{\alpha}{1+\alpha}$. 
In~\eqref{eq:m12_M22_inv:c}, we use $\underline{1}^\transpose_{N_\text{o}}\bOmega_1 = \left(\frac{1+2\alpha}{1+\alpha}+\sum_{i=1}^{N_\text{o}-1}\frac{\alpha}{\alpha+1}\right)\underline{1}_{N_\text{o}}^\transpose = \frac{\alpha N_\text{o}+\alpha+1}{\alpha+1}\underline{1}_{N_\text{o}}^\transpose$, which follows from summing the diagonal and off-diagonal contributions of $\bOmega_1$.
%
%

\begin{figure}[h]
    \centering
        \begin{subfigure}[t]{1.0\columnwidth}
        \centering
        \includegraphics[width=\columnwidth]{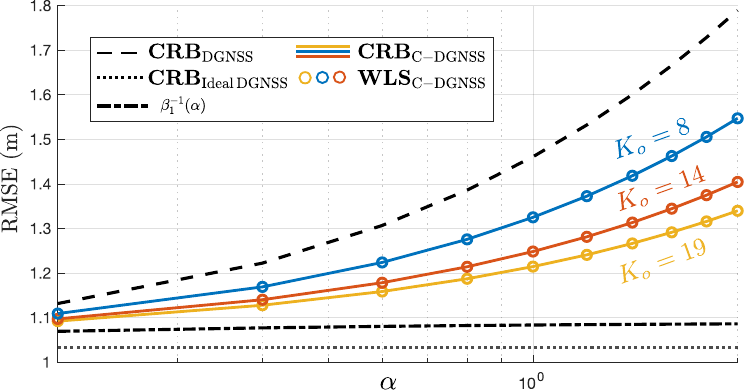}
        \caption{Impact of the inter-user correlation factor $\alpha$ for $N_o=10$.}
        \label{fig:results_vs_alpha}
    \end{subfigure}
    
    \begin{subfigure}[t]{1.0\columnwidth}
        \centering
        \includegraphics[width=\columnwidth]{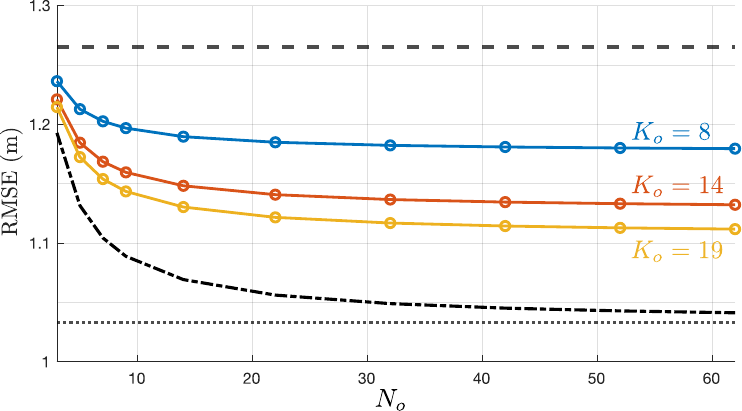}
        \caption{Impact of the number of aiding users $N_o$ for $\alpha = 0.5$.}
        \label{fig:results_vs_No}
    \end{subfigure}
\caption{Impact of cooperative network parameters on 3D positioning RMSE in C-DGNSS. Results are shown for $K_o = \{8,\,14,\,19\}$. We report $\mathbf{CRB}_{\mathrm{DGNSS}}$~\eqref{eq:bound:noncoop}, $\mathbf{CRB}_{\mathrm{Ideal\,DGNSS}}$~\eqref{eq:bound:ideal}, the asymptotic benchmark $\beta_1(\alpha)^{-1}$~\eqref{eq:proof_1}, and, for the C-DGNSS model in Table~\ref{tab:crb_model_summary}, $\mathbf{CRB}_{\mathrm{C\text{-}DGNSS}}$~\eqref{eq:bound_cdgnss} and $\mathbf{WLS}_{\mathrm{C\text{-}DGNSS}}$~\eqref{eq:cdgnss_estimator}.}
\vspace{-.65cm}
\label{fig:results_cdgnss}
\end{figure}

Substituting~\eqref{eq:m12_M22_inv:c} and the $\bM_{11}$ and $\bM_{21}$ blocks from~\eqref{eq:inverse_bound} into~\eqref{eq:homogeneous_visibility_bound}, we obtain:
\begin{subequations}\label{eq:crb_float_simplified}
\begin{align}
\left. \mathbf{CRB}(\hat{\bomega}_c) \right|_{\mathbf{J}_\text{o}=0}
&=
\left(
\beta_1(\alpha)\,\mathbf{J}_\text{c}
+
\frac{\alpha\,\beta_2(\alpha)}{\alpha+1}\,
\bOmega_2
\right)^{-1}
\label{eq:crb_float_simplified:a}
\\[0.6ex]
&=
\left(
\beta_1(\alpha)
+
\frac{\alpha\,\beta_2(\alpha)}{\alpha+1}\,N_\text{o}
\right)^{-1}
\mathbf{J}_\text{c}^{-1}
\label{eq:crb_float_simplified:b}
\\[0.6ex]
&=
(1+\alpha)\,\mathbf{J}_\text{c}^{-1}
= \mathbf{CRB}_\text{Non-Coop}.
\label{eq:crb_float_simplified:c}
\end{align}
\end{subequations}
In~\eqref{eq:crb_float_simplified:a} we use the auxiliary matrix $\bOmega_2 = (\underline{1}_{N_o}^\top \otimes \bI_{K_c})(\underline{1}_{N_o}\otimes \mathbf{J}_c) = (\underline{1}_{N_o}^\top \underline{1}_{N_o}) \otimes \mathbf{J}_c = N_\text{o}\bJ_\text{c}$.

%
%
The result in~\eqref{eq:crb_float_simplified:c} shows that, for finite $N_\text{o}$, cooperation provides no gain when aiding users offer no additional visibility. In this scenario, adding users with identical information, effectively corresponding to $N_\text{c}>1$, leaves the performance unchanged.


\subsubsection{Disjoint Inter-Cluster Satellite Visibility}
We consider the case in which the constrained user shares no satellites with the aiding users, whose observation matrix in~\eqref{eq:obs_aiding} reduces to $\mathbf{G} = \mathbf{G}_\text{o}$, corresponding to observations of $K_\text{o}$ satellites only.
Under this condition, the covariance in~\eqref{eq:sigma_clusters} becomes block diagonal, with all cross-covariance terms vanishing, and its structure comprising an upper-left block $(1+\alpha)\mathbf{R}_c$ and a lower-right block $\mathbf{C}_{N_o}^{\alpha} \otimes \mathbf{R}_o$.
The FIM for this case is: 
\begin{equation*}\label{eq:FIM_disjoint}
\begin{split}
\left. \bJ(\boldsymbol{\omega}) \right|_{\mathbf{G} = \mathbf{G}_\text{o}}
\small=
\left[
\begin{array}{c@{\hspace{1em}}c}
(1+\alpha)\,\mathbf{J}_c &
\mathbf{0}_{M_\omega,\,M_\omega N_o}\\
\mathbf{0}_{M_\omega N_o,\,M_\omega}
&
\begin{bmatrix}
\beta_4(\alpha)\mathbf{J}_o & \cdots & \beta_5(\alpha)\mathbf{J}_o\\
\vdots & \ddots & \vdots\\
\beta_5(\alpha)\mathbf{J}_o & \cdots & \beta_4(\alpha)\mathbf{J}_o
\end{bmatrix}
\end{array}
\right].
\end{split}
\end{equation*}
Comparing this result with the non-disjoint case in~\eqref{eq:inverse_bound}, the aiding-user block preserves its cooperative structure, whereas the constrained-user block reduces to $(1+\alpha)\mathbf{J}_c$ rather than $\beta_1(\alpha)\mathbf{J}_c$. Consequently, the Schur complement yields the non-cooperative bound in~\eqref{eq:bound:noncoop}, indicating that disjoint satellite visibility provides no benefit to the constrained user.

\subsubsection{Large-Scale Cooperative Network} 
%
%
To characterize the asymptotic behavior of cooperation, we consider the regime where the number of aiding users grows without bound. In this limit, $\beta_{1}(\alpha) \xrightarrow[N_\text{o} \to \infty]{} 1$ and $\beta_{2}(\alpha) \xrightarrow[N_\text{o} \to \infty]{} 0$, so the \gls{FIM} in~\eqref{eq:inverse_bound} converges to
%
\begin{equation}
  \bJ(\bomega)  
\xrightarrow[N_\text{o} \to \infty]{}
\begin{bmatrix}
\mathbf{J}_\text{c} & \mathbf{0}_{M_\omega,M_\omega N_\text{o}} \\
\mathbf{0}_{M_\omega N_\text{o},M_\omega } &  \mathbf{I}_{N_\text{o}} \otimes (\mathbf{J}_\text{c}+\mathbf{J}_\text{o} )
\end{bmatrix}.
\end{equation}
The cross-information terms vanish, and the Schur complement reduces to
\begin{equation}
\mathbf{CRB}(\hat{\bomega}_c) \xrightarrow[N_\text{o} \to \infty]{} \bJ_\text{c}^{-1} = \mathbf{CRB}_\text{Ideal}.
\end{equation}
Hence, in large-scale networks, cooperation asymptotically removes the impact of the base-station noise and drives the constrained-user bound to the ideal limit determined solely by $\mathbf{J}_\text{c}$.

%

\begin{figure*}[h]
    \centering
        \includegraphics[width=.9\textwidth]{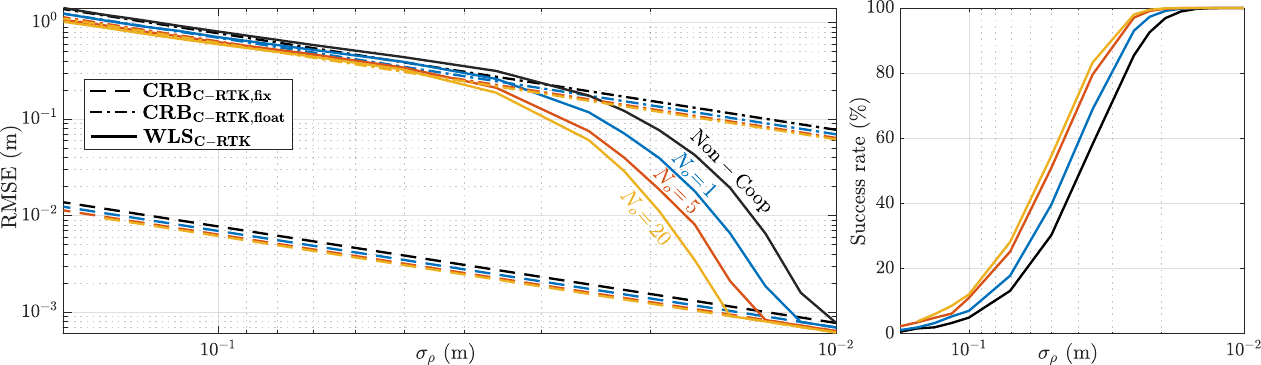}
\caption{Impact of cooperative network parameters on 3D positioning RMSE and ambiguity resolution in C-RTK. Results are reported for \( N_{o} \in \{1, 5, 25\} \).
We show (left) the bounds \( \mathbf{CRB}_{\mathrm{C\text{-}RTK,float}} \) in~\eqref{eq:crb_float}
and \( \mathbf{CRB}_{\mathrm{C\text{-}RTK,fix}} \) in~\eqref{eq:crb_freq}, together with the empirical
\( \mathbf{WLS}_{\mathrm{C\text{-}RTK}} \) performance, and (right) the success probability
\( \hat{P}_{\mathrm{succ}} \) in~\eqref{eq:success_rate}. For reference, the corresponding
non-cooperative baselines are included in all cases.
}
\vspace{-0.5cm}
\label{fig:results_crtk}
\end{figure*}

\vspace{-0.2cm}

\begin{table*}[t]
\centering
\caption{Summary of the impact of key system parameters on \gls{C-DGNSS} and C-RTK performance, as established by the theoretical performance limits in Sec.~\ref{sec:bounds} and validated via the simulation study in Sec.~\ref{sec:simulations}.}
\begin{tabular}{@{\extracolsep{\fill}}p{7cm} p{8.5cm}}
\toprule
\textbf{Parameter} & \textbf{Impact on Cooperative DGNSS/RTK} \\
\midrule
\(N_\text{o}\): Number of aiding users 
& Performance improvement for $N_\text{o}\geq 1$ subject to $K_\text{o}\geq K_\text{c}$. \\

\(K_\text{o}\): Number of satellites exclusively visible to aiding users 
& Performance improvement for $K_\text{o}\geq K_\text{c}$, subject to $N_\text{o}\geq 1$.  \\

\(\alpha\): Base--receiver variance ratio 
& Rate of improvement offered by cooperation increases with $\alpha$. \\

$\bJ_\text{c}$: FIM of the aided user (intrinsic geometric limit) 
& Performance cannot improve beyond $\bJ_\text{c}^{-1}$ even with ideal cooperation. \\

\(N_\text{c}\): Number of constrained users 
& No impact. \\

\(K_\text{c}\): Number of satellites jointly visible to aided/aiding users 
& Improved intrinsic geometric limit $\bJ_\text{c}$.  \\
\bottomrule
\end{tabular}
\label{tab:parameter_impact}
\end{table*}

\section{Simulation Experiments}\label{sec:simulations}

This section evaluates the proposed methodologies through simulation experiments.
We first outline the experimental setup, followed by an analysis of the impact of $N_\text{o}$, $K_\text{o}$, and $\alpha$ on the 3D position \gls{RMSE} and ambiguity fix success rate of the proposed cooperative differential GNSS frameworks. For a consolidated overview of the observed trends, see Table~\ref{tab:parameter_impact}.

%

%
%
%
An elevation mask of \(20^\circ\) is applied, with the base station tracking a total of \(K = 19\) satellites. The constrained visibility cluster comprises \(N_\text{c} = 2\) aided users, each observing \(K_\text{c} = 4\) satellites with \gls{GDOP} of 2.5.
%
%
%
%
%
The weighting matrix is set to $\mathbf{W}_r = \mathbf{I}$, reflecting the assumption of identical noise variance across all satellites.
To obtain statistically meaningful results, $M=10^4$ Monte Carlo runs are conducted for each parameter variation and experiment.
%
%
The performance of the cooperative differential GNSS estimator is reported for the first user in the constrained-visibility cluster and is compared against two reference limits: the model CRB and the asymptotic benchmark $\beta_1(\alpha)^{-1}$ corresponding to the ideal-visibility regime discussed in Sec.~\ref{sec:bounds}-\ref{sec:bounds:analysis}.\ref{sec:bounds:analysis:ideal}.


%
%


%
%
%

Figure~\ref{fig:results_cdgnss} highlights the complementary roles of the inter-user correlation factor $\alpha$ and the number of aiding users $N_o$ in cooperative positioning.
For increasing values of $\alpha$, the RMSE grows as expected, since a larger $\alpha$ implies a higher noise contribution from the base station, which directly inflates the effective measurement covariance of all single-differenced observations. Nevertheless, for any fixed $\alpha$, cooperation consistently improves positioning performance. The gain increases with both the number of aiding users $N_o$ and the number of additional satellites they observe $K_o$, since both factors enhance the network visibility.
For fixed $N_o$ and $\alpha$, as $K_o$ grows, the cooperative RMSE approaches the asymptotic benchmark $\beta_1^{-1}$, reflecting the ideal-visibility regime.
This trend is observed across all values of $\alpha$, confirming that sufficient cooperative visibility can compensate for increased base-station noise.

We provide additional experiments considering the C-RTK framework to assess the impact of cooperation on carrier-phase positioning and integer ambiguity resolution.
In this experiment, the base station and aiding receivers track a total of $K=K_o=8$ satellites, while the aided user ($N_c=1$) observes $K_c=4$ satellites, yielding a standalone GDOP of $2.97$. The base--rover variance ratio is set to $\alpha=1$ for both code and carrier-phase observations, and $\sigma_\Phi = \sigma_\rho/100$.
The empirical ambiguity--resolution success rate is defined as
\vspace{-0.1cm}
\begin{equation}\label{eq:success_rate}
\hat{P}_{\mathrm{succ}}
= \frac{1}{M} \sum_{m=1}^{M}
\delta\!\left(
\hat{\mathbf a}^{(m)}_{c} - \mathbf a_{\mathrm{ref},c}
\right),
\end{equation}
where $\hat{\mathbf a}^{(m)}_{c}$ denotes the estimated ambiguity vector of the constrained user at the $m$-th Monte Carlo iteration, $\mathbf a_{\mathrm{ref},c}$ is the corresponding true ambiguity vector, and $\delta(\mathbf{x})$ denotes the Kronecker delta.


%
Figure~\ref{fig:results_crtk} (left) depicts the positioning RMSE as a function of the code variance of the receivers (both aided and aiding) for different numbers of cooperating users.
Two regimes are clearly observed, namely the float and fix regimes, and in both cases cooperation yields a consistent improvement over the non-cooperative baseline. 
%
%
It can be observed that increasing the number of cooperating users shifts the transition to the fixed-ambiguity regime toward higher noise levels, indicating that cooperation enables faster and more reliable ambiguity resolution.
This effect is further quantified in Fig.~\ref{fig:results_crtk} (right), which shows a marked increase in the empirical ambiguity resolution success rate as the number of cooperating users grows.

%


\section{Conclusion}\label{sec:conclusion}









%
%
We introduce a unified estimation framework for cooperative differential GNSS that integrates code- and carrier-phase techniques while accommodating reference stations of arbitrary noise quality. Using analytical manipulations of the Fisher Information Matrix, we characterize the asymptotic performance of C-DGNSS and C-RTK as a function of receiver count, satellite visibility, and measurement variance.
Cooperation improves the positioning accuracy of users with limited satellite visibility, and the influence of base-station noise diminishes as the network size increases when cooperative users bring enhanced visibility. In large networks, the achievable accuracy converges to the limit imposed by the user’s own satellite geometry. For carrier-phase positioning, the C-RTK formulation further enhances ambiguity resolution relative to independent RTK solutions.
Simulation results validate the theoretical insights presented in this paper. Our findings show that even low-cost reference stations can deliver high-quality positioning when leveraging multiple users, offering a pathway toward more accessible differential GNSS services.

\vspace{-.25cm}
\bibliographystyle{ieeebib}
\bibliography{biblio_gnss}

\begin{IEEEbiography}[{\includegraphics[width=1in,height=1.25in,clip,keepaspectratio]{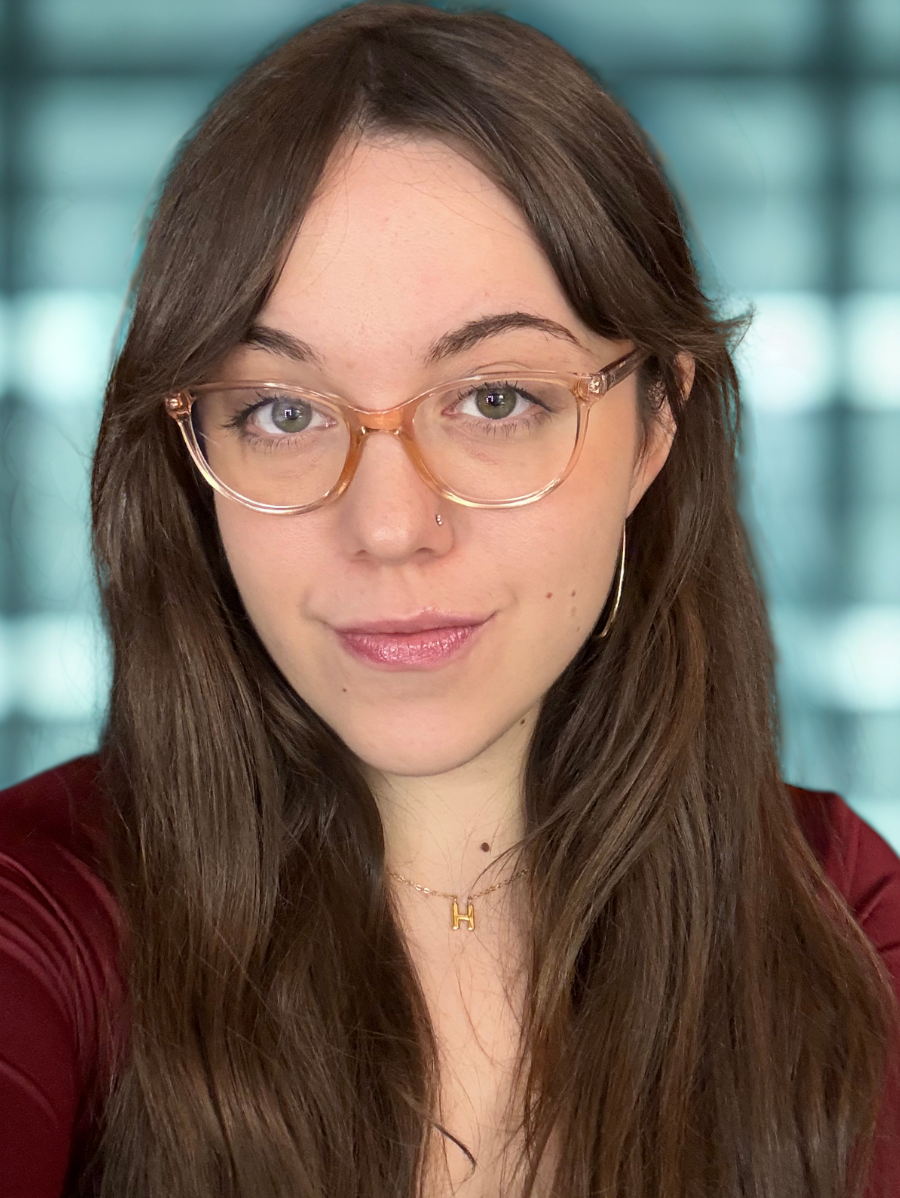}}]{Helena Calatrava}(Student Member, IEEE) received the B.S. and M.S. degrees in Electrical Engineering from UPC in 2020 and 2022, respectively. She is currently a Ph.D. candidate in Electrical and Computer Engineering at Northeastern University, Boston, MA. Her research focuses on statistical and robust signal processing, machine learning, and multitarget tracking algorithms to improve resilience in satellite-based navigation and radar-based localization systems, with an emphasis on cooperative and distributed architectures.
\end{IEEEbiography}%

\begin{IEEEbiography}[{\includegraphics[width=1in,clip,keepaspectratio]{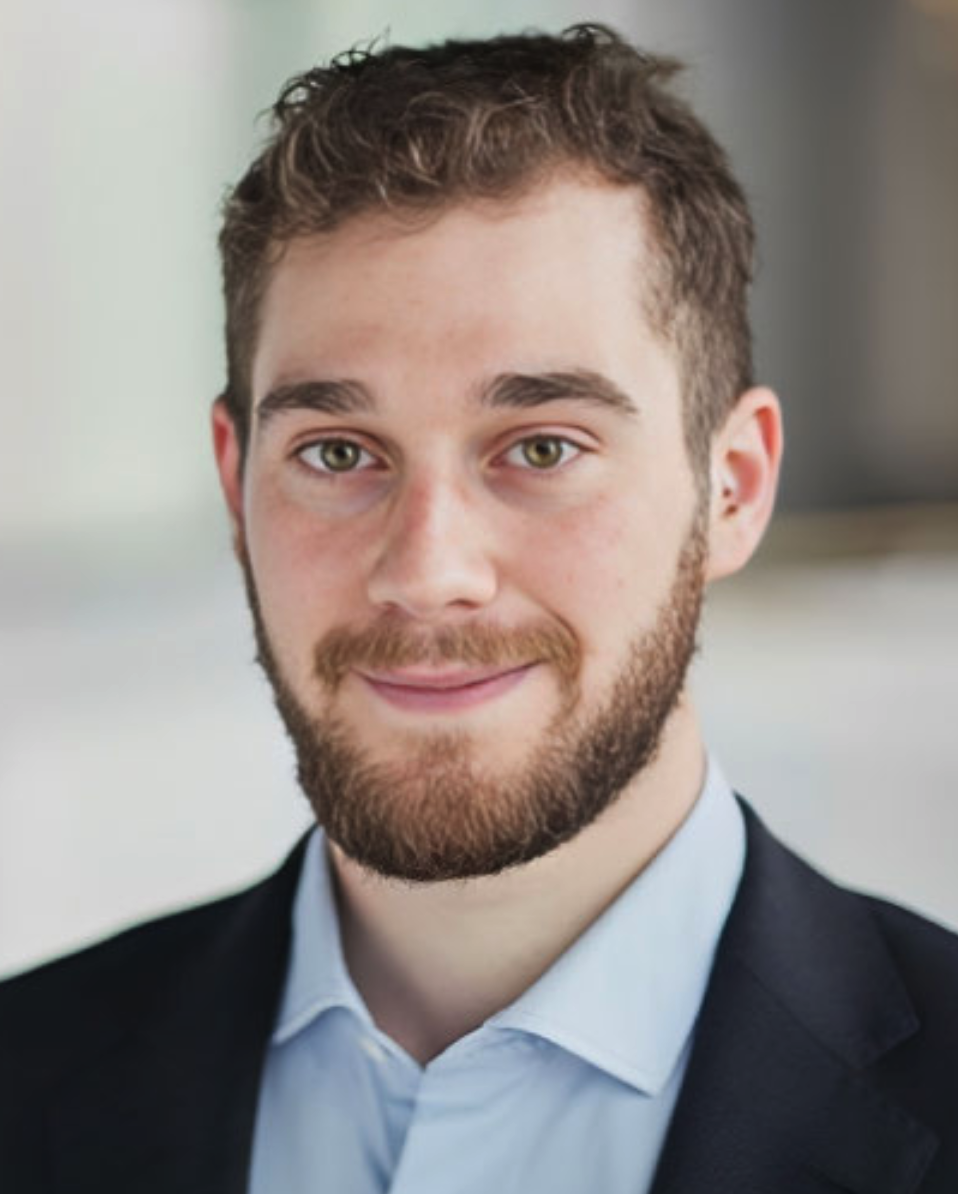}}]{Daniel Medina}{\space} received the M.S. and Ph.D. degrees in Computer Science from Universidad Carlos III de Madrid (UC3M), in 2016 and 2022, respectively. He is currently a Research Fellow with the Institute of Communications and Navigation, German Aerospace Center (DLR), where he leads the Multi-Sensor Systems Research Group. His research interests include statistical signal processing, robust filtering and estimation, LiDAR perception, and high-precision satellite-based navigation. He currently serves as Vice-Chair of the IEEE ITSS German Chapter.
\end{IEEEbiography}%

\begin{IEEEbiography}[{\includegraphics[width=1in,height=1.25in,clip,keepaspectratio]{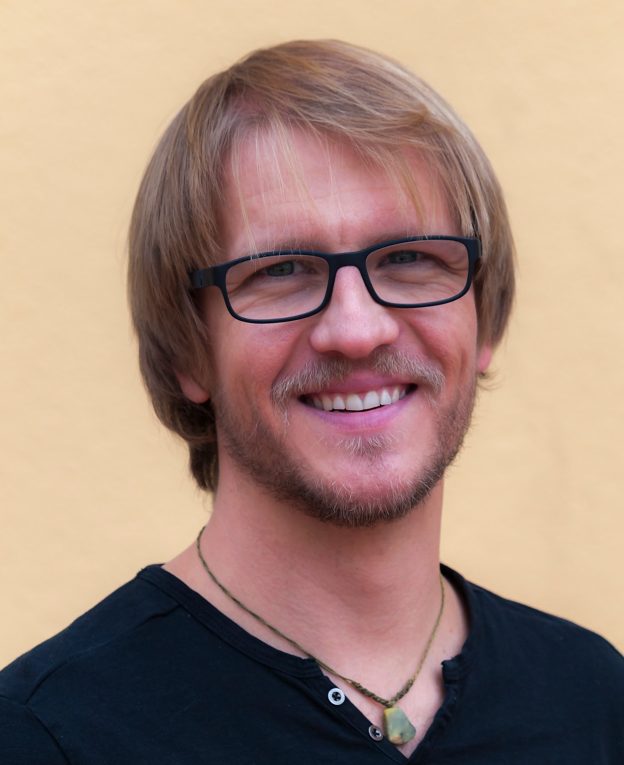}}]{Pau Closas}(Senior Member, IEEE),
is an Associate Professor in Electrical and Computer Engineering at Northeastern University, Boston MA.
He received the MS and PhD in Electrical Engineering from UPC in 2003 and 2009, respectively, and a MS in Advanced Mathematics from UPC in 2014. 
His primary areas of interest include statistical signal processing and machine learning, with applications to positioning and localization systems. 
\end{IEEEbiography}

\clearpage
\onecolumn
\setcounter{equation}{0}
\setcounter{section}{0}
\renewcommand{\theequation}{S\arabic{equation}}
\renewcommand{\thesection}{S\Roman{section}}
\renewcommand{\thesubsection}{S\Roman{section}.\Alph{subsection}}
\renewcommand{\thesubsubsection}{S\Roman{section}.\Alph{subsection}.\arabic{subsubsection}}

\begin{center}
{\Large\bfseries Supplemental Material for\\[0.3em]
``Cooperative Differential GNSS Positioning: Estimators and Bounds''\par}
\vspace{0.8em}
{\normalsize Helena Calatrava, Daniel Medina, Pau Closas\par}
\vspace{1.2em}
\end{center}

\renewcommand{\theequation}{S\arabic{equation}}
\renewcommand{\thesection}{S\Roman{section}}
\renewcommand{\thesubsection}{S\Roman{section}.\Alph{subsection}}
\renewcommand{\thesubsubsection}{S\Roman{section}.\Alph{subsection}.\arabic{subsubsection}}




Note that sections, equations, and tables in this supplement are prefixed with ``S''. References without this prefix correspond to the main paper. Notation follows as in the main paper.

Recall that, in this work, we assume a network composed of one surveyed, and potentially low-cost, reference station (indexed by~$b$) and $N$ user receivers (indexed by~$r \in \{1,\ldots,N\}$), as illustrated in Fig.~\ref{fig:intro_cdgnss}.
The reference station observes a set of $K$ satellites (indexed by~$s \in \{p,1,\ldots,K{-}1\}$), where $p$ denotes the pivot satellite used for double differencing, and each user receiver is assumed to observe all or a subset of these satellites depending on visibility conditions.

Throughout these derivations, we make use of the following Kronecker-product
identities, which enable convenient manipulation of matrices whose dimensions
grow with the cooperative nature of the algorithm:
\begin{itemize}
    \item Transpose identity: $(A \otimes B)^\top = A^\top \otimes B^\top$.
    \item Inverse identity (for invertible $A$ and $B$): $(A \otimes B)^{-1} = A^{-1} \otimes B^{-1}$.
    \item Mixed-product property: $(A \otimes B)(C \otimes D) = (AC) \otimes (BD)$,
          whenever the products $AC$ and $BD$ are well defined.
\end{itemize}

\section{Expanded Form of the Observation and Covariance Matrices}

\subsection{Single-User Model}

Recall the single-user observation model defined in~\eqref{eq:model:approximation_single}:
\begin{equation}
\begin{split}
\Delta\boldsymbol{\rho}_r &= \bH_r\,\delta\bx_r + c\,d\bt + \bm{T}_r + \bm{I}_r + \boldsymbol{\varepsilon}_r \in \mathbb{R}^K,\\
\Delta\boldsymbol{\Phi}_r &= \bH_r\,\delta\bx_r + c\,d\bt + \bm{T}_r - \bm{I}_r + \lambda\bN_r + \boldsymbol{\epsilon}_r \in \mathbb{R}^K.
\end{split}
\end{equation}
We expand the observation matrix hereafter for the sake of subsequent derivations (recall that we assume $\bH_r = \bH,\ \forall\,r$):
\begin{equation}
\mathbf{H} = [\bE,\ \b1_K]
=
\begin{bmatrix}
-(\mathbf{e}^{1})^\top & 1 \\
-(\mathbf{e}^{2})^\top & 1 \\
\vdots & \vdots \\
-(\mathbf{e}^{K})^\top & 1
\end{bmatrix}
=
\begin{bmatrix}
\underbrace{
\begin{matrix}
- e_x^{1} & - e_y^{1} & - e_z^{1} \\
- e_x^{2} & - e_y^{2} & - e_z^{2} \\
\vdots    & \vdots    & \vdots    \\
- e_x^{K} & - e_y^{K} & - e_z^{K}
\end{matrix}
}_{\mathbf{E}}
\quad
\begin{matrix}
1 \\ 1 \\ \vdots \\ 1
\end{matrix}
\end{bmatrix}\in \mathbb{R}^{K\times4},
\end{equation}
where $\be^s = -(\bp^s - \bp_{0}) / \|\bp^s - \bp_{0}\| \in \mathbb{R}^3$ is the satellite--receiver line-of-sight (LOS) unit vector.
For each user $r$, the individual code and carrier-phase error covariances are
\begin{equation}
    \boldsymbol{\Sigma}_{\rho,r} = \sigma_\rho^2\,\mathbf{W}_r^{-1} =  \begin{bmatrix}
(\sigma_{\rho,r}^{p})^{2} & & & \\
& (\sigma_{\rho,r}^{1})^{2} & & \\
& & \ddots & \\
& & & (\sigma_{\rho,r}^{K-1})^{2}
\end{bmatrix},\ \quad \boldsymbol{\Sigma}_{\Phi,r} = \sigma_\Phi^2\,\mathbf{W}_r^{-1} = \begin{bmatrix}
(\sigma_{\Phi,r}^{p})^{2} & & & \\
& (\sigma_{\Phi,r}^{1})^{2} & & \\
& & \ddots & \\
& & & (\sigma_{\Phi,r}^{K-1})^{2}
\end{bmatrix},
\end{equation}
where $\sigma_\rho^2$ and $\sigma_\Phi^2$ denote the common code and phase
variances, and $\mathbf{W}_r$ is the weighting matrix (accounting for elevation).
The superscript $(\cdot)^{k}$ indicates the measurement-specific variance for the $k$-th satellite, which depends on its elevation angle.

\subsection{Multi-User Model}

The network-stacked quantities, based on the operator in~\eqref{eq:coop_op_vector}, are as follows:
\begin{equation}
\label{eq:stacked_vectors}
\Delta\tilde{\boldsymbol{\rho}}
\triangleq
\begin{bmatrix}
\Delta\rho_b\\[0.3ex]
\Delta\rho_1\\
\vdots\\
\Delta\rho_N
\end{bmatrix},\quad
\Delta\tilde{\boldsymbol{\Phi}}
\triangleq
\begin{bmatrix}
\Delta\Phi_b\\[0.3ex]
\Delta\Phi_1\\
\vdots\\
\Delta\Phi_N
\end{bmatrix},\quad
\tilde{\boldsymbol{\varepsilon}}
\triangleq
\begin{bmatrix}
\varepsilon_b\\[0.3ex]
\varepsilon_1\\
\vdots\\
\varepsilon_N
\end{bmatrix}
\;\in\; \mathbb{R}^{K(N+1)},
\qquad
\delta\tilde{\mathbf{x}}
\triangleq
\begin{bmatrix}
\delta\mathbf{x}_b\\[0.3ex]
\delta\mathbf{x}_1\\
\vdots\\
\delta\mathbf{x}_N
\end{bmatrix}
\in \mathbb{R}^{4(N+1)}.
\end{equation}

Recall that $
\tilde{\bSigma}_\rho
= \mathbb{E}\{\tilde{\boldsymbol{\varepsilon}}\tilde{\boldsymbol{\varepsilon}}^\transpose\}=
\mathrm{diag}\!\left(
\bSigma_{\rho,b},\,
\bSigma_{\rho,1},\dots,\bSigma_{\rho,N}
\right)
\in\mathbb{R}^{K(N+1)\times K(N+1)}$. An analogous expression holds for $\tilde{\boldsymbol{\Sigma}}_\Phi$.

\subsection{C-DGNSS Model}

\textbf{$\bullet$ Observation matrix:} The single-difference (SD) operator applied across the network has the Kronecker form
\begin{equation}
\tilde{\mathbf{D}}_b
=\left[-\b1_N \otimes \mathbf{I}_K,\ \mathbf{I}_N \otimes \mathbf{I}_K\right]
=
\begin{bmatrix}
  -\mathbf{I}_K & \mathbf{I}_K & 0          & \cdots & 0 \\
  -\mathbf{I}_K & 0          & \mathbf{I}_K & \cdots & 0 \\
  \vdots        & \vdots     & \ddots       & \ddots & \vdots \\
  -\mathbf{I}_K & 0          & \cdots       & 0      & \mathbf{I}_K
\end{bmatrix} \in\mathbb{R}^{KN\times K(N{+}1)},
\end{equation}
where it is assumed for convenience that the base measurements occupy the first block in the stacked measurement vector.

Applying \(\tilde{\mathbf{D}}_b\) to the stacked pseudorange differences yields the C-DGNSS observation model as
\begin{align}\label{eq:sd_Obs}
\tilde{\mathbf{D}}_b \, \Delta\tilde{\boldsymbol{\rho}}
&=
\begin{bmatrix}
\Delta\boldsymbol{\rho}_1 - \Delta\boldsymbol{\rho}_b \\
\vdots \\
\Delta\boldsymbol{\rho}_N - \Delta\boldsymbol{\rho}_b
\end{bmatrix}
= \begin{bmatrix}
      \bH(\delta\mathbf{x}_1 - \delta\mathbf{x}_b) \\
      \vdots \\
      \bH(\delta\mathbf{x}_N - \delta\mathbf{x}_b)
    \end{bmatrix} + \tilde{\boldsymbol{\varepsilon}}_b
\\
&= \begin{bmatrix}
 \mathbf{H} &  &  \\
  & \ddots &    \\
 &  & \mathbf{H}
\end{bmatrix}\begin{bmatrix}
\delta\mathbf{x}_1 - \delta\mathbf{x}_b \\
\vdots \\
\delta\mathbf{x}_N - \delta\mathbf{x}_b
\end{bmatrix} + \tilde{\boldsymbol{\varepsilon}}_b
=
\boxed{
    \tilde{\mathbf{H}}_N \, \delta\tilde{\mathbf{x}}_b
    + \tilde{\boldsymbol{\varepsilon}}_b
    \triangleq
    \Delta\tilde{\boldsymbol{\rho}}_b \in\mathbb{R}^{KN}}.
\end{align}

\textbf{$\bullet$ Covariance matrix:} The C-DGNSS covariance matrix can be calculated as
\begin{align}
\tilde{\boldsymbol{\Sigma}}_{b,\rho} &= \mathbb{E}\{\tilde{\boldsymbol{\varepsilon}}_b\, \tilde{\boldsymbol{\varepsilon}}_b^\transpose\}
=
\tilde{\mathbf{D}}_b\,\tilde{\boldsymbol{\Sigma}}_\rho\,\tilde{\mathbf{D}}_b^\top
\\
&=
\begin{cases}
\bSigma_{\rho,i} + \bSigma_{\rho,b}, & i=j,\\[1ex]
\bSigma_{\rho,b}, & i\neq j.
\end{cases}
\end{align}

\begin{equation}
\boxed{
\tilde{\boldsymbol{\Sigma}}_{b,\rho}
=
\begin{bmatrix}
\bSigma_{\rho,1}+\bSigma_{\rho,b} & \bSigma_{\rho,b} & \cdots & \bSigma_{\rho,b} \\
\bSigma_{\rho,b} & \bSigma_{\rho,2}+\bSigma_{\rho,b} & \cdots & \bSigma_{\rho,b} \\
\vdots & \vdots & \ddots & \vdots \\
\bSigma_{\rho,b} & \bSigma_{\rho,b} & \cdots & \bSigma_{\rho,N}+\bSigma_{\rho,b}
\end{bmatrix}
}
\end{equation}
Here, the diagonal blocks capture intra-user terms, with diagonal entries reflecting the combined base--user variance for each satellite and zero off-diagonals. The off-diagonal blocks contain only base-station noise: their diagonals include the base variance for each satellite, and their off-diagonals are zero, indicating no inter-user cross-correlation beyond the shared base.

\textbf{$\bullet$ Other quantities:} The following expanded expressions make explicit the stacking, differencing, and matrix structure underlying the compact model in~(13).

\begin{equation}\label{eq:explicit_sd}
\centering
\begin{aligned}
\Delta\tilde{\boldsymbol{\rho}}_b
&\triangleq
\begin{bmatrix}
\Delta\boldsymbol{\rho}_{b1} \\
\vdots \\
\Delta\boldsymbol{\rho}_{bN}
\end{bmatrix}
=
\begin{bmatrix}
\Delta\boldsymbol{\rho}_1 - \Delta\boldsymbol{\rho}_b \\
\vdots \\
\Delta\boldsymbol{\rho}_N - \Delta\boldsymbol{\rho}_b
\end{bmatrix}
\in \mathbb{R}^{KN},\quad
\Delta\tilde{\boldsymbol{\Phi}}_b
\triangleq
\begin{bmatrix}
\Delta\boldsymbol{\Phi}_{b1} \\
\vdots \\
\Delta\boldsymbol{\Phi}_{bN}
\end{bmatrix}
=
\begin{bmatrix}
\Delta\boldsymbol{\Phi}_1 - \Delta\boldsymbol{\Phi}_b \\
\vdots \\
\Delta\boldsymbol{\Phi}_N - \Delta\boldsymbol{\Phi}_b
\end{bmatrix}
\in \mathbb{R}^{KN},
\\[1.1ex]
\delta\tilde{\mathbf{x}}_b
&\triangleq
\begin{bmatrix}
\delta\mathbf{x}_{b1} \\
\vdots \\
\delta\mathbf{x}_{bN}
\end{bmatrix}
=
\begin{bmatrix}
\delta\mathbf{x}_1 - \delta\mathbf{x}_b \\
\vdots \\
\delta\mathbf{x}_N - \delta\mathbf{x}_b
\end{bmatrix}
\in \mathbb{R}^{4N},
\tilde{\boldsymbol{\varepsilon}}_b
\triangleq
\begin{bmatrix}
\tilde{\boldsymbol{\varepsilon}}_{b1} \\
\vdots \\
\tilde{\boldsymbol{\varepsilon}}_{bN}
\end{bmatrix}
=
\begin{bmatrix}
\boldsymbol{\varepsilon}_1 - \boldsymbol{\varepsilon}_b \\
\vdots \\
\boldsymbol{\varepsilon}_N - \boldsymbol{\varepsilon}_b
\end{bmatrix}
\in \mathbb{R}^{KN},
\tilde{\boldsymbol{\epsilon}}_b
\triangleq
\begin{bmatrix}
\tilde{\boldsymbol{\epsilon}}_{b1} \\
\vdots \\
\tilde{\boldsymbol{\epsilon}}_{bN}
\end{bmatrix}
=
\begin{bmatrix}
\boldsymbol{\epsilon}_1 - \boldsymbol{\epsilon}_b \\
\vdots \\
\boldsymbol{\epsilon}_N - \boldsymbol{\epsilon}_b
\end{bmatrix}
\in \mathbb{R}^{KN}.
\end{aligned}
\end{equation}

\subsection{C-RTK Model}

\textbf{$\bullet$ Observation matrix:} The double-difference (DD) operator applied across the network has the form

\begin{equation}
\mathbf{D}_p =[\, -\mathbf{1}_{K-1},\ \mathbf{I}_{K-1} \,] 
=
\begin{bmatrix}
-1 & 1 & 0 & \cdots & 0 \\
-1 & 0 & 1 & \cdots & 0 \\
\vdots & \vdots & \vdots & \ddots & \vdots \\
-1 & 0 & 0 & \cdots & 1
\end{bmatrix}
\in\mathbb{R}^{(K-1)\times K},
\quad
\tilde{\mathbf{D}}_p
=\bI_N\ \otimes\ \bD_p
=
\begin{bmatrix}
\mathbf{D}_p & \mathbf{0}   & \cdots & \mathbf{0} \\
\mathbf{0}   & \mathbf{D}_p & \cdots & \mathbf{0} \\
\vdots       & \vdots       & \ddots & \vdots \\
\mathbf{0}   & \mathbf{0}   & \cdots & \mathbf{D}_p
\end{bmatrix}\in\mathbb{R}^{(K-1)N\times KN},
\end{equation}
where it is assumed for convenience that the pivot measurements occupy the first block in the stacked measurement vector.

Applying the pivot-differencing operator to the stacked SD
observations in~\eqref{eq:sd_Obs} gives
\begin{equation}\label{eq:dd_Obs}
\tilde{\mathbf{D}}_p \,\Delta\tilde{\brho}_b
= \underbrace{( \mathbf{I}_N \otimes \mathbf{D}_p ) 
  ( \mathbf{I}_N \otimes \mathbf{H} )}_{\mathbf{I}_N \otimes (\mathbf{D}_p \mathbf{H})}
  \,\delta\tilde{\mathbf{x}}_b + \tilde{\bvarepsilon}_b =  
  \underbrace{\begin{bmatrix}
\mathbf{D}_p\mathbf{H}\,\delta\mathbf{x}_{b1} \\
\vdots \\
\mathbf{D}_p\mathbf{H}\,\delta\mathbf{x}_{bN}
\end{bmatrix}}_{\eqref{aux}}
+ \tilde{\bvarepsilon}_b = \boxed{\tilde{\bD}_p\tilde{\mathbf{E}}_N\tilde{\bb} + \tilde{\boldsymbol{\varepsilon}}_b^p  \triangleq \Delta\brho_b^{p} \in \mathbb{R}^{(K-1)N}}.
\end{equation}
Note that here we use
\begin{equation}\label{aux}
  \begin{bmatrix}
\mathbf{D}_p\mathbf{H}\,\delta\mathbf{x}_{b1} \\
\vdots \\
\mathbf{D}_p\mathbf{H}\,\delta\mathbf{x}_{bN}
\end{bmatrix} = 
    \begin{bmatrix}
 \mathbf{D}_p &  &  \\
  & \ddots &    \\
 &  & \mathbf{D}_p
\end{bmatrix}\begin{bmatrix}
 \mathbf{H} &  &  \\
  & \ddots &    \\
 &  & \mathbf{H}
\end{bmatrix}\delta\tilde{\mathbf{x}}_b
= 
    \begin{bmatrix}
 \mathbf{D}_p &  &  \\
  & \ddots &    \\
 &  & \mathbf{D}_p
\end{bmatrix}\begin{bmatrix}
 \mathbf{E} &  &  \\
  & \ddots &    \\
 &  & \mathbf{E}
\end{bmatrix}\delta\tilde{\mathbf{p}}_b,
\end{equation}
%
where the equivalence in the second equality can be understood from:
\begin{equation}
\mathbf{D}_p \mathbf{H} \delta\mathbf{x}_{br}
= 
\underbrace{\begin{bmatrix}
(\be^{1}-\be^{p})^\top & 0 \\
(\be^{2}-\be^{p})^\top & 0 \\
\vdots & \vdots \\
(\be^{K}-\be^{p})^\top & 0
\end{bmatrix}}_{\in\ \mathbb{R}^{(K-1)\times 4}} \begin{bmatrix}
\mathbf{p}_r - \mathbf{p}_b \\
c\,(dt_r - dt_b)
\end{bmatrix}
=
\underbrace{\begin{bmatrix}
(\mathbf{e}^{1}-\mathbf{e}^{p})^\top \\
(\mathbf{e}^{2}-\mathbf{e}^{p})^\top \\
\vdots \\
(\mathbf{e}^{K}-\mathbf{e}^{p})^\top
\end{bmatrix}}_{\in\mathbb{R}^{(K-1)\times 3}}
\left(\mathbf{p}_r - \mathbf{p}_b\right) = \mathbf{D}_p \mathbf{E} \delta\mathbf{p}_{br}.
\end{equation}
Furthermore, note that we use $\tilde{\bb} = \delta\tilde{\mathbf{p}}_{b}$ to refer to the full-network baseline vector (between the base station $b$ and each receiver).

\textbf{$\bullet$ Covariance matrix:} The C-RTK covariance matrix can be calculated as

\begin{align}
 \tilde{\bD}_p\,\tilde{\boldsymbol{\Sigma}}_{b,\rho}\,\tilde{\bD}_p^\transpose &= \begin{bmatrix}
 \mathbf{D}_p &  &  \\
  & \ddots &    \\
 &  & \mathbf{D}_p
\end{bmatrix} \begin{bmatrix}
\bSigma_{\rho,1}+\bSigma_{\rho,b} & \bSigma_{\rho,b} & \cdots & \bSigma_{\rho,b} \\
\bSigma_{\rho,b} & \bSigma_{\rho,2}+\bSigma_{\rho,b} & \cdots & \bSigma_{\rho,b} \\
\vdots & \vdots & \ddots & \vdots \\
\bSigma_{\rho,b} & \bSigma_{\rho,b} & \cdots & \bSigma_{\rho,N}+\bSigma_{\rho,b}
\end{bmatrix} \begin{bmatrix}
 \mathbf{D}_p^\transpose &  &  \\
  & \ddots &    \\
 &  & \mathbf{D}_p^\transpose
\end{bmatrix}
\\
&=
%
\boxed{
\begin{cases}
\mathbf{D}_p(\bSigma_{\rho,i}+\bSigma_{\rho,b})\mathbf{D}_p^\top,
& i=j, \\[0.6em]
\mathbf{D}_p\bSigma_{\rho,b}\mathbf{D}_p^\top,
& i\neq j.
\end{cases}
\quad \triangleq\ \tilde{\bSigma}_{b,\rho}^{p}}.
\end{align}
An analogous expression holds for $\tilde{\bSigma}_{b,\Phi}^{p}$.
From this expression, we study how the noise contributions from the base station and each user appear within the block-diagonal covariance structure for the C-RTK model (see Fig.~\ref{fig:covariance_motivation}).

\textbf{Diagonal blocks of $\tilde{\bSigma}_{b,\rho}^{p}$ $(i=j)$:}

\begin{align}
    \mathbf{D}_p(\bSigma_{\rho,i}+\bSigma_{\rho,b})\mathbf{D}_p^\top &= \mathbf{D}_p\begin{bmatrix}
(\sigma_{\rho,i}^{p})^{2} + (\sigma_{\rho,b}^{p})^{2}& & & \\
& (\sigma_{\rho,i}^{1})^{2} +(\sigma_{\rho,b}^{1})^{2} & & \\
& & \ddots & \\
& & & (\sigma_{\rho,i}^{K-1})^{2} + (\sigma_{\rho,b}^{K-1})^{2}
\end{bmatrix} \mathbf{D}_p^\transpose 
\\
&= 
\begin{bmatrix}
-(\sigma_{\rho,i}^{p})^{2} - (\sigma_{\rho,b}^{p})^{2}
    & (\sigma_{\rho,i}^{1})^{2} + (\sigma_{\rho,b}^{1})^{2}
    & 
    & 
    &  \\[0.8ex]
\vdots
    & 
    & \ddots
    &
    &  \\[0.8ex]
-(\sigma_{\rho,i}^{p})^{2} - (\sigma_{\rho,b}^{p})^{2}
    & 
    & 
    & 
    & (\sigma_{\rho,i}^{K-1})^{2} + (\sigma_{\rho,b}^{K-1})^{2}
\end{bmatrix} \begin{bmatrix}
-1 & \cdots & -1 \\[0.6ex]
1  &        &        \\[0.6ex]
   & \ddots &        \\[0.6ex]
   &        & 1
\end{bmatrix}
\\
&=
\begin{cases}
(\sigma_{\rho,i}^{p})^{2} + (\sigma_{\rho,b}^{p})^{2}
+ (\sigma_{\rho,i}^{i})^{2} + (\sigma_{\rho,b}^{i})^{2}, & i = j, \\[1ex]
(\sigma_{\rho,i}^{p})^{2} + (\sigma_{\rho,b}^{p})^{2}, & i \neq j.
\end{cases}
\end{align}
The diagonal blocks capture the intra-user correlation terms: their diagonal entries contain a fourfold variance contribution (from both the base and the user at the pivot satellite, and from both the base and the user at the corresponding satellite), and the off-diagonal elements contain a twofold contribution arising solely from the pivot-satellite terms (from base and user).

\textbf{Off-diagonal blocks of $\tilde{\bSigma}_{b,\rho}^{p}$ $(i\neq j)$:}
\begin{align}
    \mathbf{D}_p\bSigma_{\rho,b}\mathbf{D}_p^\top &= \mathbf{D}_p\begin{bmatrix}
(\sigma_{\rho,b}^{p})^{2}& & & \\
& (\sigma_{\rho,b}^{1})^{2} & & \\
& & \ddots & \\
& & &  (\sigma_{\rho,b}^{K-1})^{2}
\end{bmatrix} \mathbf{D}_p^\transpose 
=
\begin{cases}
(\sigma_{\rho,b}^{p})^{2} +  (\sigma_{\rho,b}^{i})^{2}, & i = j, \\[1ex]
 (\sigma_{\rho,b}^{p})^{2}, & i \neq j.
\end{cases}
\end{align}
The off-diagonal blocks encode the inter-user correlation terms and contain only contributions from the base-station noise. Within each block, the diagonal entries exhibit a twofold contribution from the base station (one term associated with the common satellite and one with the pivot), whereas the off-diagonal entries include only the pivot-satellite contribution.

\section{Full Derivation of the Parameterized Fisher Information Matrix}\label{sec:s1}

In this section, we present the complete derivations leading to $\tilde{\bR}^{-1}$ in~\eqref{eq:full_sigma_inv_aligned} and to the Fisher information matrix (FIM) $\mathbf{J}(\boldsymbol{\tilde{\omega}})$ in~\eqref{eq:inverse_bound}, as referenced in Sec.~\ref{sec:bounds}--\ref{sec:bounds:fisher}. The expressions of the coefficients $\beta_i(\alpha)$, $i \in \{0,1,\dots,5\}$ listed in Table~\ref{tab:beta_coefficients} are also derived.
Importantly, the network clustering model presented in Sec.~\ref{sec:bounds}-\ref{sec:network_clustering} must be considered hereafter.

\subsection{Covariance Matrix Inverse}

We first derive the expression of $\tilde{\bR}^{-1}$ for arbitrary $N_\text{o}$ and $\alpha$ via the Schur complement of~\eqref{eq:sigma_clusters}, where we use $\bX = (1+\alpha)\bR_\text{c}$, $\bY = \bP^\alpha_{N_\text{o}} = \b1_{N_{\mathrm{o}}}^\top \otimes 
\begin{bmatrix}
\alpha\mathbf{R}_{\mathrm{c}} & \mathbf{0}_{K_{\mathrm{c}},\,K_{\mathrm{o}}}
\end{bmatrix}$, and $\bZ = \tilde{\bC}^\alpha_{N_\text{o}}\otimes\bR$, and denote the four blocks of $\tilde{\bR}^{-1}$ by $[\tilde{\bR}^{-1}]_{ij}$ for $i,j\in\{1,2\}$.

\textbf{$\bullet$ Step 1 (top-diagonal block):}
To obtain $[\tilde{\bR}^{-1}]_{11} 
= (\bX - \bY \bZ^{-1} \bY^\top)^{-1} $, let us start by calculating $\bZ^{-1}$ as:
%
%

    %

\begin{equation}
\begin{split}
    &\bZ^{-1} = 
    (\tilde{\mathbf{C}}_{N_\text{o}}^\alpha\otimes\bR)^{-1} 
    %
    =
    [\mathbf{I}_{N_\text{o}} + \alpha\mathbf{1}_{N_\text{o}}]^{-1}\otimes\bR^{-1}
    \overset{\mathrm{}}{=}  \underbrace{\left(\bI_{N_\text{o}}-\frac{\alpha}{\alpha N_\text{o}+1}\mathbf{1}_{N_\text{o}}\right)}_{\boldsymbol{\Gamma}_1}\otimes\ \bR^{-1},
    %
    %
\end{split}
\end{equation}
The diagonal and off-diagonal elements of the auxiliary matrix $\bGamma_1$ are $\frac{1+\alpha(N_\text{o}-1)}{\alpha N_\text{o}+1}$ and $-\frac{\alpha}{\alpha N_\text{o}+1}$, respectively. In the last equality, we have used the Sherman--Morrison formula, which allows the inversion of a rank-one modification of the identity matrix as 
\begin{equation}\label{eq:sherman_morrison}
    (\mathbf{I} + \mathbf{u}\mathbf{v}^\top)^{-1} = \mathbf{I} - \frac{\mathbf{u}\mathbf{v}^\top}{1 + \mathbf{v}^\top \mathbf{u}},
\end{equation}
with $\mathbf{u} = \mathbf{v} = \sqrt{\alpha}\b1_{N_\text{o}}$.  
Now we left-multiply by matrix $\bY = \bP^\alpha_{N_\text{o}} = \left(\b1_{N_\text{o}}^\transpose \otimes 
\begin{bmatrix}
\alpha\bR_\text{c} & \mathbf{0}_{K_\text{c}, K_\text{o}}
\end{bmatrix}\right)$:
\begin{equation}\label{eq:s3}
\begin{split}
    \bY \bZ^{-1}  &=  
%
\left(\b1_{N_\text{o}}^\transpose \otimes 
\begin{bmatrix}
\alpha\bR_\text{c} & \mathbf{0}_{K_\text{c}, K_\text{o}}
\end{bmatrix}\right) \left(\bGamma_1\otimes \bR^{-1}\right) = 
\left(\b1_{N_\text{o}}^\transpose 
\bGamma_1\right)\otimes  \left( \begin{bmatrix}
\alpha\bR_\text{c} & \mathbf{0}_{K_\text{c}, K_\text{o}}
\end{bmatrix} \underbrace{\begin{bmatrix}
\bR_\text{c}^{-1} & \mathbf{0}  \\
\mathbf{0} &  \bR_\text{o}^{-1}
\end{bmatrix}}_{\bR^{-1}}\right)
\\&=\left(\frac1{\alpha N_\text{o}+1}\b1_{N_\text{o}}^\transpose\right)\otimes   \begin{bmatrix}
\bR_\text{c}\bR_\text{c}^{-1} & \mathbf{0}_{K_\text{c}, K_\text{o}}
\end{bmatrix} = 
\alpha\underbrace{\frac1{\alpha N_\text{o}+1}}_{\beta_0(\alpha)} \left(\b1_{N_\text{o}}^\transpose \otimes 
\begin{bmatrix}
\mathbf{I}_{K_\text{c}} & \mathbf{0}_{K_\text{c}, K_\text{o}}
\end{bmatrix}\right),
\end{split}
\end{equation}
where we have used $\b1^\transpose_{N_\text{o}}\bGamma_1 = \left(\frac{1+\alpha(N_\text{o}-1)}{\alpha N_\text{o}+1}-\sum_{i=1}^{N_\text{o}-1}\frac{\alpha}{\alpha N_\text{o}+1}\right)\b1_{N_\text{o}}^\transpose = \frac1{\alpha N_\text{o}+1}\b1_{N_\text{o}}^\transpose$, which follows from summing the diagonal and off-diagonal contributions of $\bGamma_1$.
Now we can calculate
%
\begin{equation}
\begin{split}
   [\tilde{\bR}^{-1}]_{11} 
&= (\bX - \bY \bZ^{-1} \bY^\top)^{-1} =
\left((1+\alpha)\bR_\text{c} - \alpha\beta_0(\alpha)
\left(\b1_{N_\text{o}}^\transpose \otimes 
\begin{bmatrix}
\mathbf{I}_{K_\text{c}} & \mathbf{0}_{K_\text{c}, K_\text{o}}
\end{bmatrix}\right)
\left(\b1_{N_\text{o}} \otimes \begin{bmatrix}
\alpha\bR_\text{c}\\ \mathbf{0}_{K_\text{o}, K_\text{c}}
\end{bmatrix}\right)
\right)^{-1}
\\
&=
\left((1+\alpha)\bR_\text{c} - \alpha\beta_0(\alpha)
\underbrace{(\b1_{N_\text{o}}^\transpose\b1_{N_\text{o}})}_{N_\text{o}}\otimes
\underbrace{\left(
\begin{bmatrix}
\mathbf{I}_{K_\text{c}} & \mathbf{0}_{K_\text{c}, K_\text{o}}
\end{bmatrix}\begin{bmatrix}
\alpha\bR_\text{c}\\ \mathbf{0}_{K_\text{o}, K_\text{c}}
\end{bmatrix}\right)}_{\alpha\bR_\text{c}} \right)^{-1}
=
\left((1+\alpha)\bR_\text{c} - \alpha^2\beta_0(\alpha)
N_\text{o}\bR_\text{c} \right)^{-1}
\\&= (1+\alpha -\alpha^2\beta_0(\alpha) N_\text{o})^{-1}\bR_\text{c}^{-1}= \underbrace{\frac{\alpha N_\text{o} + 1}{\alpha N_\text{o} + \alpha + 1}}_{\beta_1(\alpha)}\bR_\text{c}^{-1}.
\end{split}
\end{equation}
%
%
%

%
\textbf{$\bullet$ Step 2 (off-diagonal blocks):} The blocks $[\tilde{\bR}^{-1}]_{12}$ and $[\tilde{\bR}^{-1}]_{21}$ are calculated as:
\begin{equation}
\begin{split}
[\tilde{\bR}^{-1}]_{12} &= [\tilde{\bR}^{-1}]_{21}^\top
= -[\tilde{\bR}^{-1}]_{11} \, \bY \bZ^{-1} = -\alpha\beta_0(\alpha)\beta_1(\alpha)\bR_\text{c}^{-1} \left(\b1_{N_\text{o}}^\transpose \otimes 
\begin{bmatrix}\bI_{K_\text{c}} & \mathbf{0}_{K_\text{c}, K_\text{o}}
\end{bmatrix}\right) = \underbrace{-\frac{\alpha}{\alpha N_\text{o} + \alpha + 1}}_{\beta_2(\alpha)} \left(\b1_{N_\text{o}}^\transpose \otimes 
\begin{bmatrix}\bR_\text{c}^{-1} & \mathbf{0}_{K_\text{c}, K_\text{o}}
\end{bmatrix}\right).
\end{split}
\end{equation}
%

\textbf{$\bullet$ Step 3 (bottom-diagonal block): } To calculate $[\tilde{\bR}^{-1}]_{22} = \bZ^{-1} - \bZ^{-1} \bY^\top \, [\tilde{\bR}^{-1}]_{12}$, we start calculating the term
%
%
\begin{equation}
\begin{split}
\bZ^{-1} \bY^\top \, [\tilde{\bR}^{-1}]_{12} &=  \underbrace{\alpha\beta_0(\alpha) \beta_2(\alpha)}_{\beta_3(\alpha)}\left(\b1_{N_\text{o}}\otimes\begin{bmatrix}
\bI_{K_\text{c}}\\ \mathbf{0}_{K_\text{o}, K_\text{c}}
\end{bmatrix}\right)\left(\b1_{N_\text{o}}^\transpose \otimes 
\begin{bmatrix}\bR_\text{c}^{-1} & \mathbf{0}_{K_\text{c}, K_\text{o}}
\end{bmatrix}\right)
= \beta_3(\alpha)\underbrace{\left(\b1_{N_\text{o}}\b1_{N_\text{o}}^\transpose\right)}_{\mathbf{1}_{N_\text{o}}}\otimes\begin{bmatrix}
    \bR_\text{c}^{-1}&\mathbf{0}_{K_\text{c},K_\text{o}}\\\mathbf{0}_{K_\text{o},K_\text{c}}&\mathbf{0}_{K_\text{o}}
\end{bmatrix}
%
%
%
%
%
\\
&=\beta_3(\alpha)
\begin{bmatrix}
\begin{bmatrix}
\bR_\text{c}^{-1} & \mathbf{0}_{K_\text{c}, K_\text{o}} \\
\mathbf{0}_{K_\text{o}, K_\text{c}} & \mathbf{0}_{K_\text{o}}
\end{bmatrix}
&
\cdots
&
\begin{bmatrix}
\bR_\text{c}^{-1} & \mathbf{0}_{K_\text{c}, K_\text{o}} \\
\mathbf{0}_{K_\text{o}, K_\text{c}} & \mathbf{0}_{K_\text{o}}
\end{bmatrix}
\\
\vdots & \ddots & \vdots \\
\begin{bmatrix}
\bR_\text{c}^{-1} & \mathbf{0}_{K_\text{c}, K_\text{o}} \\
\mathbf{0}_{K_\text{o}, K_\text{c}} & \mathbf{0}_{K_\text{o}}
\end{bmatrix}
&
\cdots
&
\begin{bmatrix}
\bR_\text{c}^{-1} & \mathbf{0}_{K_\text{c}, K_\text{o}} \\
\mathbf{0}_{K_\text{o}, K_\text{c}} & \mathbf{0}_{K_\text{o}}
\end{bmatrix}
\end{bmatrix},
\end{split}
\end{equation}
and now we can calculate
%
\begin{equation}
\begin{split}
[\tilde{\bR}^{-1}]_{22}
&= \bZ^{-1} - \bZ^{-1} \bY^\top \, [\tilde{\bR}^{-1}]_{12} 
=
\boldsymbol{\Gamma}_1\otimes\ \bR^{-1} - 
\beta_3(\alpha)\mathbf{1}_{N_\text{o}}\otimes\begin{bmatrix}
    \bR_\text{c}^{-1}&\mathbf{0}_{K_\text{c},K_\text{o}}\\\mathbf{0}_{K_\text{o},K_\text{c}}&\mathbf{0}_{K_\text{o}}
\end{bmatrix} 
\\&=
\begin{bmatrix}
\underbrace{\dfrac{1+\alpha (N_\text{o}-1)}{\alpha N_\text{o} + 1}}_{\beta_4(\alpha)}\,\mathbf{R}^{-1} 
& \cdots 
& \underbrace{-\dfrac{\alpha}{\alpha N_\text{o} + 1}}_{\beta_5(\alpha)}\,\mathbf{R}^{-1}
\\[1.2ex]
\vdots & \ddots & \vdots
\\[1.2ex]
-\dfrac{\alpha}{\alpha N_\text{o} + 1}\,\mathbf{R}^{-1} 
& \cdots 
& \dfrac{1+\alpha (N_\text{o}-1)}{\alpha N_\text{o} + 1}\,\mathbf{R}^{-1}
\end{bmatrix} - 
\beta_3(\alpha)\begin{bmatrix}
\begin{bmatrix}
\bR_\text{c}^{-1} & \mathbf{0}_{K_\text{c}, K_\text{o}} \\
\mathbf{0}_{K_\text{o}, K_\text{c}} & \mathbf{0}_{K_\text{o}}
\end{bmatrix}
&
\cdots
&
\begin{bmatrix}
\bR_\text{c}^{-1} & \mathbf{0}_{K_\text{c}, K_\text{o}} \\
\mathbf{0}_{K_\text{o}, K_\text{c}} & \mathbf{0}_{K_\text{o}}
\end{bmatrix}
\\
\vdots & \ddots & \vdots \\
\begin{bmatrix}
\bR_\text{c}^{-1} & \mathbf{0}_{K_\text{c}, K_\text{o}} \\
\mathbf{0}_{K_\text{o}, K_\text{c}} & \mathbf{0}_{K_\text{o}}
\end{bmatrix}
&
\cdots
&
\begin{bmatrix}
\bR_\text{c}^{-1} & \mathbf{0}_{K_\text{c}, K_\text{o}} \\
\mathbf{0}_{K_\text{o}, K_\text{c}} & \mathbf{0}_{K_\text{o}}
\end{bmatrix}
\end{bmatrix}
\\&=
\begin{bmatrix}
\begin{bmatrix}
\beta_{3,4}(\alpha)\bR_\text{c}^{-1} & \mathbf{0}_{K_\text{c}, K_\text{o}} \\
\mathbf{0}_{K_\text{o}, K_\text{c}} & \beta_4(\alpha)\bR_\text{o}^{-1}
\end{bmatrix}
&
\cdots
&
\begin{bmatrix}
\beta_{3,5}(\alpha)\bR_\text{c}^{-1} & \mathbf{0}_{K_\text{c}, K_\text{o}} \\
\mathbf{0}_{K_\text{o}, K_\text{c}} & \beta_5(\alpha)\bR_\text{o}^{-1}
\end{bmatrix}
\\
\vdots & \ddots & \vdots \\
\begin{bmatrix}
\beta_{3,5}(\alpha)\bR_\text{c}^{-1} & \mathbf{0}_{K_\text{c}, K_\text{o}} \\
\mathbf{0}_{K_\text{o}, K_\text{c}} & \beta_5(\alpha)\bR_\text{o}^{-1}
\end{bmatrix}
&
\cdots
&
\begin{bmatrix}
\beta_{3,4}(\alpha)\bR_\text{c}^{-1} & \mathbf{0}_{K_\text{c}, K_\text{o}} \\
\mathbf{0}_{K_\text{o}, K_\text{c}} & \beta_4(\alpha)\bR_\text{o}^{-1}
\end{bmatrix}
\end{bmatrix}
\\&=
\bI_N \otimes \bF_1 + (\mathbf{1}_N - \bI_N) \otimes \bF_2.
\end{split}
\end{equation}
%
%
%
Here, $\beta_{i,j}(\alpha) = \beta_{j}(\alpha) -\beta_i(\alpha)$. Note that $\beta_{3,4}(\alpha) = \beta_1(\alpha)$ and $\beta_{3,5}(\alpha) = \beta_2(\alpha)$. Additionally, in the last equality, we have reduced to a more compact form by using the following matrices:
\begin{equation}
\begin{split}
    \bF_{m} &=
    \beta_{m+3}(\alpha)\begin{bmatrix}
        \bR^{-1}_\text{c}\ \mathbf{0}_{K_\text{c},K_\text{o}} \\\mathbf{0}_{K_\text{o},K_\text{c}}\ \bR^{-1}_\text{o} 
    \end{bmatrix} -
    \beta_3(\alpha)
    \begin{bmatrix}\bR_\text{c}^{-1}\ \mathbf{0}_{K_\text{c},K_\text{o}}\\\mathbf{0}_{K_\text{o},K_\text{c}}\ \mathbf{0}_{K_\text{o}}
\end{bmatrix}
    = 
    \begin{bmatrix}
        \beta_{3,m+3}(\alpha)\bR^{-1}_\text{c} &\mathbf{0}_{K_\text{c},K_\text{o}} \\\mathbf{0}_{K_\text{o},K_\text{c}} &\beta_{m+3}(\alpha)\bR^{-1}_\text{o} 
    \end{bmatrix}\ \quad \forall\ m\in\{1,\ 2\}.
    \end{split}
\end{equation}
%
%
%

%
%
%
\textbf{$\bullet$ Step 4 (final expression):} The expanded expression for $\tilde{\bR}^{-1}$ is as follows (by putting together the blocks calculated in the previous steps):
\begin{equation}\label{eq:full_sigma_inv_aligned}
\boxed{
\tilde{\bR}^{-1} =
\left[
\begin{array}{c@{\hspace{1em}}c}
\displaystyle \underbrace{\beta_1(\alpha) \, \bR_\text{c}^{-1}}_{[\tilde{\bR}^{-1}]_{11}} &
\underbrace{\begin{bmatrix}
\begin{bmatrix}
\beta_2(\alpha) \, \bR_\text{c}^{-1} & \mathbf{0}_{K_\text{c}, K_\text{o}}
\end{bmatrix} &
\cdots &
\begin{bmatrix}
\beta_2(\alpha) \, \bR_\text{c}^{-1} & \mathbf{0}_{K_\text{c}, K_\text{o}}
\end{bmatrix}
\end{bmatrix}}_{[\tilde{\bR}^{-1}]_{12}}
\\[2ex]
\underbrace{\begin{bmatrix}
\begin{bmatrix}
\beta_2(\alpha) \, \bR_\text{c}^{-1} \\
\mathbf{0}_{K_\text{o}, K_\text{c}}
\end{bmatrix} \\
\vdots \\
\begin{bmatrix}
\beta_2(\alpha) \, \bR_\text{c}^{-1} \\
\mathbf{0}_{K_\text{o}, K_\text{c}}
\end{bmatrix}
\end{bmatrix}}_{[\tilde{\bR}^{-1}]_{21}}
&
\underbrace{\begin{bmatrix}
\begin{bmatrix}
\beta_{1}(\alpha)\bR_\text{c}^{-1} & \mathbf{0}_{K_\text{c}, K_\text{o}} \\
\mathbf{0}_{K_\text{o}, K_\text{c}} & \beta_4(\alpha)\bR_\text{o}^{-1}
\end{bmatrix}
&
\cdots
&
\begin{bmatrix}
\beta_{2}(\alpha) \bR_\text{c}^{-1} & \mathbf{0}_{K_\text{c}, K_\text{o}} \\
\mathbf{0}_{K_\text{o}, K_\text{c}} & \beta_5(\alpha)\bR_\text{o}^{-1}
\end{bmatrix}
\\
\vdots & \ddots & \vdots \\
\begin{bmatrix}
\beta_{2}(\alpha) \bR_\text{c}^{-1} & \mathbf{0}_{K_\text{c}, K_\text{o}} \\
\mathbf{0}_{K_\text{o}, K_\text{c}} & \beta_5(\alpha)\bR_\text{o}^{-1}
\end{bmatrix}
&
\cdots
&
\begin{bmatrix}
\beta_{1}(\alpha) \bR_\text{c}^{-1} & \mathbf{0}_{K_\text{c}, K_\text{o}} \\
\mathbf{0}_{K_\text{o}, K_\text{c}} & \beta_4(\alpha)\bR_\text{o}^{-1}
\end{bmatrix}
\end{bmatrix}}_{[\tilde{\bR}^{-1}]_{22}}
\end{array}
\right]
}.
\end{equation}

\begin{table}[t]
\centering
\caption{$\beta_i(\alpha)$ coefficient definitions.}
\label{tab:beta_coefficients}
\renewcommand{\arraystretch}{1.7}
\setlength{\tabcolsep}{4pt}
\begin{tabular}{@{}c c c c c c c@{}}
\toprule
 & $\beta_0$ & $\beta_1$ & $\beta_2$ & $\beta_3$ & $\beta_4$ & $\beta_5$ \\ 
\midrule

\textbf{Compact} 
& --- 
& $\displaystyle (\alpha\beta_0 + 1)^{-1}$ 
& $\displaystyle -\alpha\beta_1\beta_0$
& $\displaystyle \alpha\beta_2\beta_0$
& $\displaystyle 1 - \alpha\beta_0$
& $\displaystyle -\alpha\beta_0$
\\[1ex]

\textbf{Full} 
& $\displaystyle \frac{1}{1+\alpha N_\text{o}}$ 
& $\displaystyle \frac{\alpha N_\text{o} + 1}{\alpha N_\text{o} + \alpha + 1}$
& $\displaystyle -\frac{\alpha}{\alpha N_\text{o} + \alpha + 1}$
& $\displaystyle -\frac{\alpha^2}{(1+\alpha N_\text{o})(\alpha N_\text{o} + \alpha + 1)}$
& $\displaystyle \frac{1+\alpha(N_\text{o} - 1)}{\alpha N_\text{o} + 1}$
& $\displaystyle -\frac{\alpha}{\alpha N_\text{o} + 1}$
\\
\bottomrule
\end{tabular}
\end{table}

\subsection{Fisher Information Matrix (FIM)}
The FIM follows directly from the unified linear model in~\eqref{eq:unified_model}. Owing to the block structure of $\tilde{\bR}^{-1}$ in~\eqref{eq:full_sigma_inv_aligned}, multiplication by $\tilde{\mathbf{G}} $ preserves this structure and maps each block of $\tilde{\bR}^{-1}$ into the corresponding block of $\mathbf{J}(\boldsymbol{\omega})$. This yields the compact expression in~\eqref{eq:inverse_bound}, which is used in Sec.~\ref{sec:bounds}--\ref{sec:bounds:analysis} to analyze the asymptotic performance of the C-DGNSS/C-RTK estimators.

\begin{equation}
    \bJ(\bomega) = \tilde{\bG}^\transpose \tilde{\bR}^{-1} \tilde{\bG} = \begin{bmatrix}
\mathbf G_\text{c}^\top 
& \mathbf 0_{M_\omega,K_\text{c}} \\[0.6em]
\mathbf 0_{K_\text{c},M_\omega} 
& \mathbf I_{N_\text{o}} \otimes 
\begin{bmatrix}
\mathbf G_\text{c}^\top \;\; \mathbf G_\text{o}^\top
\end{bmatrix}
\end{bmatrix}\tilde{\bR}^{-1}\begin{bmatrix}
\mathbf G_\text{c} &  \mathbf{0}_{K_\text{c},M_\omega}   \\[0.6em]

\mathbf{0}_{M_\omega,K_\text{c}}  &\bI_{N_\text{o}}\otimes \begin{bmatrix}
\mathbf G_\text{c} \\[0.2em]
\mathbf G_\text{o}
\end{bmatrix}
\end{bmatrix}
\end{equation}

\textbf{$\bullet$ Step 1 (top-diagonal block):}
\begin{equation}
    [\bJ(\bomega)]_{11} = \bG_\text{c}^\transpose [\tilde{\bR}^{-1}]_{11} \bG_\text{c}= \beta_1(\alpha) \bG_\text{c}^\transpose \bR_\text{c}^{-1} \bG_\text{c} = \boxed{\beta_1(\alpha)\bJ_\text{c} \triangleq \bM_{11}},
\end{equation}
where $\bJ_\text{c}$ is the constrained-user FIM defined in~\eqref{eq:fims}, together with the $\bJ_\text{o}$ representing the exclusive information for aiding users.

\textbf{$\bullet$ Step 2 (off-diagonal blocks):}

\begin{equation}
    \bG^\transpose \begin{bmatrix}
\beta_2(\alpha) \, \bR_\text{c}^{-1} \\
\mathbf{0}_{K_\text{o}, K_\text{c}}
\end{bmatrix}\bG_\text{c}
= \begin{bmatrix}
\mathbf G_\text{c}^\top \;\; \mathbf G_\text{o}^\top
\end{bmatrix} \begin{bmatrix}
\beta_2(\alpha) \, \bR_\text{c}^{-1} \\
\mathbf{0}_{K_\text{o}, K_\text{c}}
\end{bmatrix}\bG_\text{c}
= 
\beta_2(\alpha)\bG_\text{c}^\transpose\bR_\text{c}^{-1}\bG_\text{c} = \beta_2(\alpha)\bJ_\text{c}.
\end{equation}

\begin{equation}
    [\bJ(\bomega)]_{21} = [\bJ(\bomega)]_{12}^\transpose = \b1_{N_\text{o}} \otimes (\beta_2(\alpha)\bJ_\text{c}) = \boxed{\beta_2(\alpha)(\b1_{N_\text{o}}\otimes\bJ_\text{c}) \triangleq \bM_{21} = \bM_{12}^\transpose}.
\end{equation}

\textbf{$\bullet$ Step 3 (bottom-diagonal blocks):} The block $\mathbf{M}_{22}$ is given by
\begin{align}
   \mathbf{M}_{22} =  
   \left(\mathbf I_{N_\text{o}} \otimes 
\bG^\transpose\right)\underbrace{\begin{bmatrix}
\begin{bmatrix}
\beta_{1}(\alpha)\bR_\text{c}^{-1} & \mathbf{0}_{K_\text{c}, K_\text{o}} \\
\mathbf{0}_{K_\text{o}, K_\text{c}} & \beta_4(\alpha)\bR_\text{o}^{-1}
\end{bmatrix}
&
\cdots
&
\begin{bmatrix}
\beta_{2}(\alpha) \bR_\text{c}^{-1} & \mathbf{0}_{K_\text{c}, K_\text{o}} \\
\mathbf{0}_{K_\text{o}, K_\text{c}} & \beta_5(\alpha)\bR_\text{o}^{-1}
\end{bmatrix}
\\
\vdots & \ddots & \vdots \\
\begin{bmatrix}
\beta_{2}(\alpha) \bR_\text{c}^{-1} & \mathbf{0}_{K_\text{c}, K_\text{o}} \\
\mathbf{0}_{K_\text{o}, K_\text{c}} & \beta_5(\alpha)\bR_\text{o}^{-1}
\end{bmatrix}
&
\cdots
&
\begin{bmatrix}
\beta_{1}(\alpha) \bR_\text{c}^{-1} & \mathbf{0}_{K_\text{c}, K_\text{o}} \\
\mathbf{0}_{K_\text{o}, K_\text{c}} & \beta_4(\alpha)\bR_\text{o}^{-1}
\end{bmatrix}
\end{bmatrix}}_{[\mathbf{R}^{-1}]_{22}}\left(\bI_{N_\text{o}}\otimes \bG\right),
\end{align}
where we can see two contributions: one from multiplying $\mathbf{G}^\top$ with the diagonal terms of $[\mathbf{R}^{-1}]_{22}$, and another from multiplying $\mathbf{G}^\top$ with its off-diagonal terms. We calculate each of these contributions and add them up, as follows:

\begin{align}
   \text{Diagonal contribution: }
   &\bG^\transpose \begin{bmatrix}
\beta_{1}(\alpha) \bR_\text{c}^{-1} & \mathbf{0}_{K_\text{c}, K_\text{o}} \\
\mathbf{0}_{K_\text{o}, K_\text{c}} & \beta_4(\alpha)\bR_\text{o}^{-1}
\end{bmatrix} \bG = 
\begin{bmatrix}
\beta_{1}(\alpha)\mathbf G_\text{c}^\top\bR_\text{c}^{-1} \;\; \beta_{4}(\alpha)\mathbf G_\text{o}^\top\bR_\text{o}^{-1} 
\end{bmatrix}\begin{bmatrix}
\mathbf G_\text{c} \\[0.2em]
\mathbf G_\text{o}
\end{bmatrix} \\&= \beta_{1}(\alpha)\mathbf G_\text{c}^\top\bR_\text{c}^{-1}\bG_\text{c} + \beta_{4}(\alpha)\mathbf G_\text{o}^\top\bR_\text{o}^{-1}\bG_\text{o} = \beta_{1}(\alpha)\bJ_\text{c} + \beta_{4}(\alpha)\bJ_\text{o},
\end{align}

\begin{align}
   \text{Off-diagonal contribution: }
   &\bG^\transpose \begin{bmatrix}
\beta_{2}(\alpha) \bR_\text{c}^{-1} & \mathbf{0}_{K_\text{c}, K_\text{o}} \\
\mathbf{0}_{K_\text{o}, K_\text{c}} & \beta_5(\alpha)\bR_\text{o}^{-1}
\end{bmatrix} \bG = 
\begin{bmatrix}
\beta_{2}(\alpha)\mathbf G_\text{c}^\top\bR_\text{c}^{-1} \;\; \beta_{5}(\alpha)\mathbf G_\text{o}^\top\bR_\text{o}^{-1} 
\end{bmatrix}\begin{bmatrix}
\mathbf G_\text{c} \\[0.2em]
\mathbf G_\text{o}
\end{bmatrix} \\&= \beta_{2}(\alpha)\mathbf G_\text{c}^\top\bR_\text{c}^{-1}\bG_\text{c} + \beta_{5}(\alpha)\mathbf G_\text{o}^\top\bR_\text{o}^{-1}\bG_\text{o} = \beta_{2}(\alpha)\bJ_\text{c} + \beta_{5}(\alpha)\bJ_\text{o}.
\end{align}
To integrate these calculations into the full block, we use 
\begin{align}
    \bM_{22} &= \bI_{N_\text{o}}\otimes(\beta_{1}(\alpha)\bJ_\text{c} + \beta_{4}(\alpha)\bJ_\text{o}) + (\mathbf{1}_{N_\text{o}} - \bI_{N_\text{o}})\otimes(\beta_{2}(\alpha)\bJ_\text{c} + \beta_{5}(\alpha)\bJ_\text{o})
    \\
    &= \boxed{\left(\bI_{N_\text{o}} + \beta_{2}(\alpha)\,\mathbf{1}_{N_\text{o}}\right)\!\otimes\bJ_\text{c} + \left(\bI_{N_\text{o}} + \beta_{5}(\alpha)\,\mathbf{1}_{N_\text{o}}\right)\!\otimes\bJ_\text{o} = \bM_{22}}.
\end{align}

\end{document}